\documentclass[a4paper,fleqn,usenatbib]{mnras}

\usepackage{newtxtext,newtxmath}

\usepackage[T1]{fontenc}
\usepackage{ae,aecompl}

\usepackage{graphicx}	
\usepackage{amsmath}	

\usepackage{subcaption}
\usepackage{xcolor}

\title[Angular-Velocity Offsets in Disk Galaxies]{On the Presence of Angular-Velocity Offsets in Disk Galaxies}

\author[Tomer Zimmerman, Lev Tal-Or \& Roy Gomel]
{
Tomer Zimmerman$^{1}$ \thanks{E-mail: stomerzi@gmail.com}
Lev Tal-Or$^{1}$$^,$$^{2}$ \thanks{E-mail: levtalor@ariel.ac.il}
Roy Gomel$^{3}$ \thanks{E-mail: roygomel6@gmail.com}
\\
$^{1}$ Department of Physics, Ariel University, Ariel 40700, Israel\\
$^{2}$ Astrophysics, Geophysics and Space Science Research Center, Ariel University, Ariel 40700, Israel\\
$^{3}$ School of Physics and Astronomy, Tel Aviv University, Ramat Aviv 69978, Israel
\\
}

\date{Accepted XXX. Received YYY; in original form ZZZ}

\pubyear{2025}

\begin{document}
\label{firstpage}
\pagerange{\pageref{firstpage}--\pageref{lastpage}}
\maketitle


\begin{abstract}
The well-known discrepancy in galactic rotation curves refers to the mismatch between observed rotational velocities and the velocities predicted by baryonic matter. In this study we investigate a potential pattern in the discrepancy, which may point to an underlying pattern in dark-halo distributions. By looking at rotational-velocity curves from an alternate perspective, the angular-velocity curves, it appears that the observed angular velocities and their corresponding baryonic predictions differ by a constant shift. That is, the discrepancy may be reduced to a constant angular-velocity term, independent of the radius. We test the generality of the suggested property by analysing 143 high-quality rotation curves. The property appears significant as it performs equally well (or better) than well-established models. Compared to a Burkert dark-halo profile, it is preferred in 60\% of the cases, relative to a Navarro–Frenk–White profile (NFW), it is superior in 73\% of the cases, and relative to Modified Newtonian Dynamics (MOND), it exhibits similar performance, being favoured in 50\% of the cases. Next, by including the new phenomenological property within the dynamical equations, we find an explicit expression for the dark-halo profile. The new single-parameter profile is characterised by an interesting property: it is intrinsically related to the baryonic distribution. Thus, information regarding the cuspy or cored nature of a particular dark halo, according to this profile, is encoded (and explicitly determined) by the respective baryonic behaviour.
 
\end{abstract}




\section{Introduction}

\begin{figure*}
\begin{subfigure}{0.3\textwidth}
    \includegraphics[width=\textwidth]{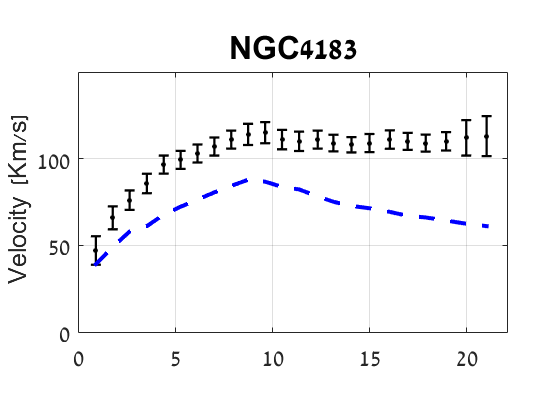}
    \vspace{-2.5\baselineskip}
\end{subfigure}
\begin{subfigure}{0.3\textwidth}
    \includegraphics[width=\textwidth]{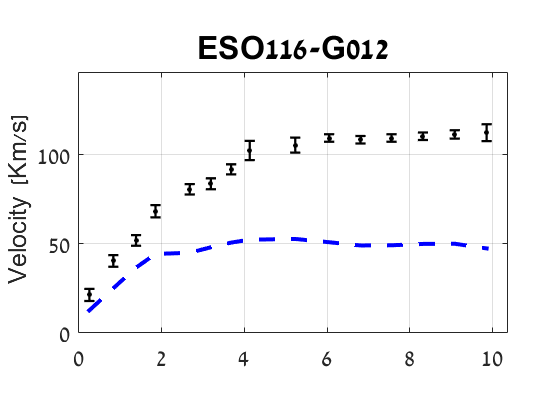}
    \vspace{-2.5\baselineskip}
\end{subfigure}
\begin{subfigure}{0.3\textwidth}
    \includegraphics[width=\textwidth]{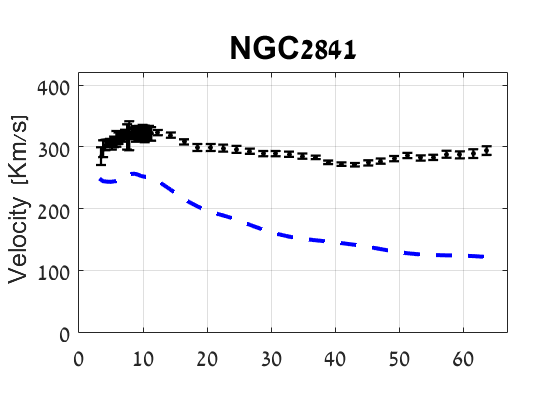}
    \vspace{-2.5\baselineskip}
\end{subfigure}
\begin{subfigure}{0.3\textwidth}
    \includegraphics[width=\textwidth]{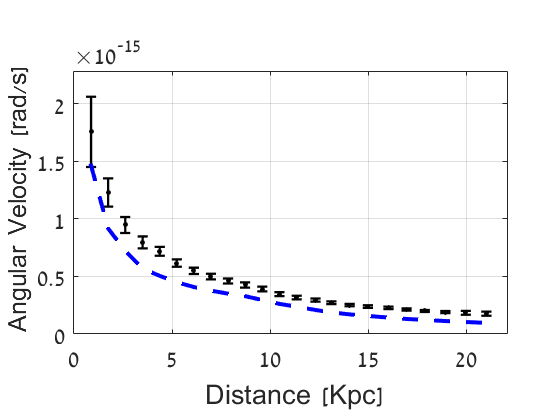}
\end{subfigure}
\begin{subfigure}{0.3\textwidth}
    \includegraphics[width=\textwidth]{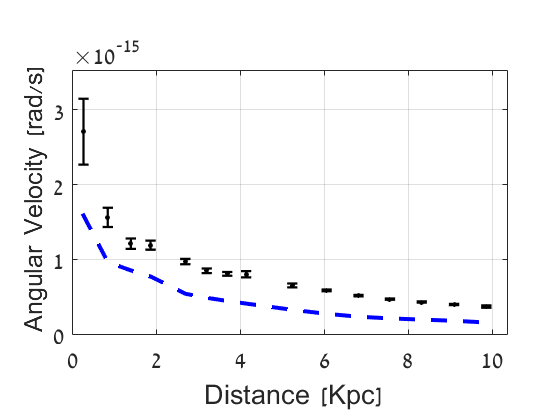}
\end{subfigure}
\begin{subfigure}{0.3\textwidth}
    \includegraphics[width=\textwidth]{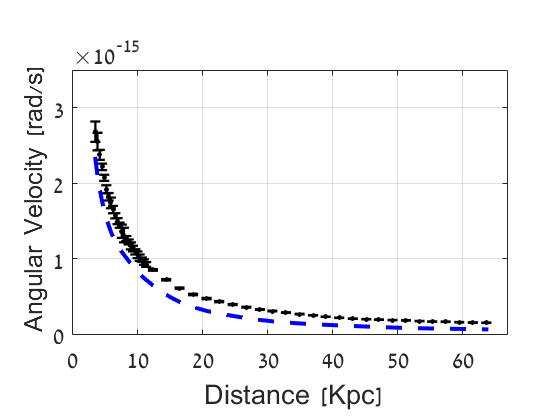}
\end{subfigure}
\caption{Rotation curves and angular-velocity curves of three different galaxies. Top panels: measured rotation curves (black points with error-bars) and their corresponding baryonic contributions (dashed blue lines). Bottom panels: measured angular-velocity curves (black points with error-bars) and their baryonic contributions (dashed blue lines). Apparently, the discrepancy in galactic rotation curves is reduced to a constant offset in this picture. The $\rm M/L$ values: NGC4183: 1, ESO116-G012: 0.5, NGC2841: 0.82. Each baryonic component includes both the stellar disk and the gas contributions.}
\label{fig:AVC_example}
\end{figure*}

One of the most useful tools to analyse the dynamics of disk galaxies and derive their mass distributions is a galactic rotation curve \citep{Bosma_1981AJ, 2001ARAA_39_137S}. A rotation curve (RC) presents the circular velocities in a galaxy as a function of the distance from the galaxy's center. Historically, the discrepancy revealed in RCs is known as the first robust evidence for dark matter \citep{Babcock1939, Salpeter1978, 1978ApJ_225L_107R, 1980ApJ_238_471R, 1985ApJS_58_107C, SancisiAlbada1987}. Illustrations of the discrepancy (i.e. the mismatch between baryonic predictions and observations) are shown in the upper panels of Fig.~\ref{fig:AVC_example}. 

Over the years, advances in observational technology have significantly enhanced the precision of RC measurements. The advent of high-resolution radio interferometers enabled detailed mapping of hydrogen gas distribution in galaxies, providing better velocity profiles in the outer regions where stars are too faint to be detected \citep{2008AJ_136_2563W, 2008AJ_136_2648D}. The development of integral field spectroscopy at optical and infrared wavelengths has allowed for even more accurate velocity measurements, enabling the study of galaxies in increased detail and at farther distances \citep{2015ApJ_798_7B, 2017AJ_154_86W_Wake, Genzel2017_mar, 2019MNRAS_485_Tiley}. 

Significant progress has also been made in baryonic modeling. Traditionally, the baryonic contribution to the rotational velocities was calculated by applying analytical mass models, most notably a Freeman exponential disk \citep{1970ApJ_Freeman}. Modern methods, though, rely on observed surface-brightness and gas profiles to model the baryonic mass distributions. These may be obtained through deep photometric observations using optical and near-infrared (NIR) telescopes. NIR imaging (e.g., using the \textit{Spitzer} Space Telescope) is preferred for tracing stellar mass because it is less affected by dust extinction \citep{1996AA_deJong, 2011_apr_McDonald, 2016AJ_Lelli_152, 2018_56_Galliano}. 

Although the baryonic distributions are less dominant than their accompanying dark halos \citep{2012_nov_PDU_Kuhlen, 2013ApJ_771_Reddick}, their impact on the overall dynamics seems to be important. A notable result of such a kind was found by \cite{Sancisi_2004}, who showed that the features in baryonic distributions closely correlate with features in rotation curves. Since the velocities in rotation curves reflect the total gravitational field (i.e., of the overall mass distribution), such a correlation highlights the important role of baryons in describing the total dynamics. 

In the present study, we describe a general property of disk galaxies of a similar nature. By switching the view to angular-velocity curves, instead of rotational velocity curves, we suggest that the velocities inferred from baryonic distributions not only correlate with observed rotation curves but also display similarities with their shapes. The discrepancy between observations and baryonic predictions, under this view, may be reduced to a constant offset.

The objective of the current paper is to demonstrate, establish, and explore the implications of the constant-offsets property. A natural outcome of the property is an explicit relation that connects dark and baryonic distributions in disk galaxies. This can be achieved by phenomenologically integrating the offsets into the dynamics. The resulting relation can be interpreted as a new data-driven dark-halo profile. 

Another notable result is the ability to obtain galaxies' baryonic masses directly from the data. By using the constant-offset property with data from the outer parts of a rotation curve, it becomes possible to compute the total baryonic mass of a galaxy without assuming how the galaxy's dark (or baryonic) components are distributed.

The potential presence of constant offsets in disk galaxies may also prompt inquiries into the underlying cause of this property. In the last section, we discuss the coupling mechanism between dark and baryonic distributions and its potential relation to the current pattern. We also demonstrate that angular-velocity shifts could be treated as a side effect arising from the rotation of the inertial frame. We mention the idea and discuss its possible implications. 

The next sections are organized as follows: Section 2 presents the new concept of angular-velocity curves and explores the universality of constant offsets. Section 3 calculates the baryonic masses of 128 disk galaxies using their observed rotation curves, and without additional assumptions. Section 4 derives a relation that connects baryonic distributions with their host dark halos, or equivalently, introduces a new dark-halo profile. In Section 5, we discuss potential interpretations of the observed property, conclude, and propose future directions.

\section{Angular-Velocity Curves}\label{sec:AVCs}

An angular-velocity curve is simply a rotation curve that presents angular velocities instead of actual velocities. At first sight it seems trivial to introduce such a curve since it provides no new information on the dynamics or kinematics of a given galaxy. However, angular-velocity curves (AVCs) are a more natural way to introduce the following property.

In Fig.~\ref{fig:AVC_example} we plot AVCs of three different galaxies. Each panel presents the observed AVC together with its expected baryonic contribution. The observed data, as well as the numerical baryonic models, were obtained from the SPARC database \citep{SPARC}. The mass-to-light values ($\rm M/L$) were taken from \cite{Li_mar_2020}. Note that the angular velocities are at the order of $\rm 10^{-15}$\,rad/s which corresponds to periods of $\sim200$\,Myr.

Looking at the figure, a curious pattern is revealed: the baryonic and the observed angular-velocity distributions seem to differ by a constant offset. That is to say that the shapes and structures of the distributions show similar behaviour.

In what follows we wish to investigate whether offsets are a general property of AVCs. However, before diving into the analysis, let us clarify that constant offsets in AVCs do not imply a flat term for the dark component (i.e. a contribution of a flat angular-velocity term from the dark halo). This is due to the fact that the angular-velocity contributions (i.e., of the dark and the baryonic components) cannot be directly summed up. This is essentially the same as in regular rotation curves, where the summation of the velocity components is in quadrature. Now, if an offset cannot be solely attributed to a dark halo, it must be regarded as a combined (yet unknown) property of the dark and baryonic distributions. Section \ref{sec:Relation} examines a possible link of this type.

\begin{figure*}
\centering
\begin{subfigure}{0.47\textwidth}
    \includegraphics[width=\textwidth]{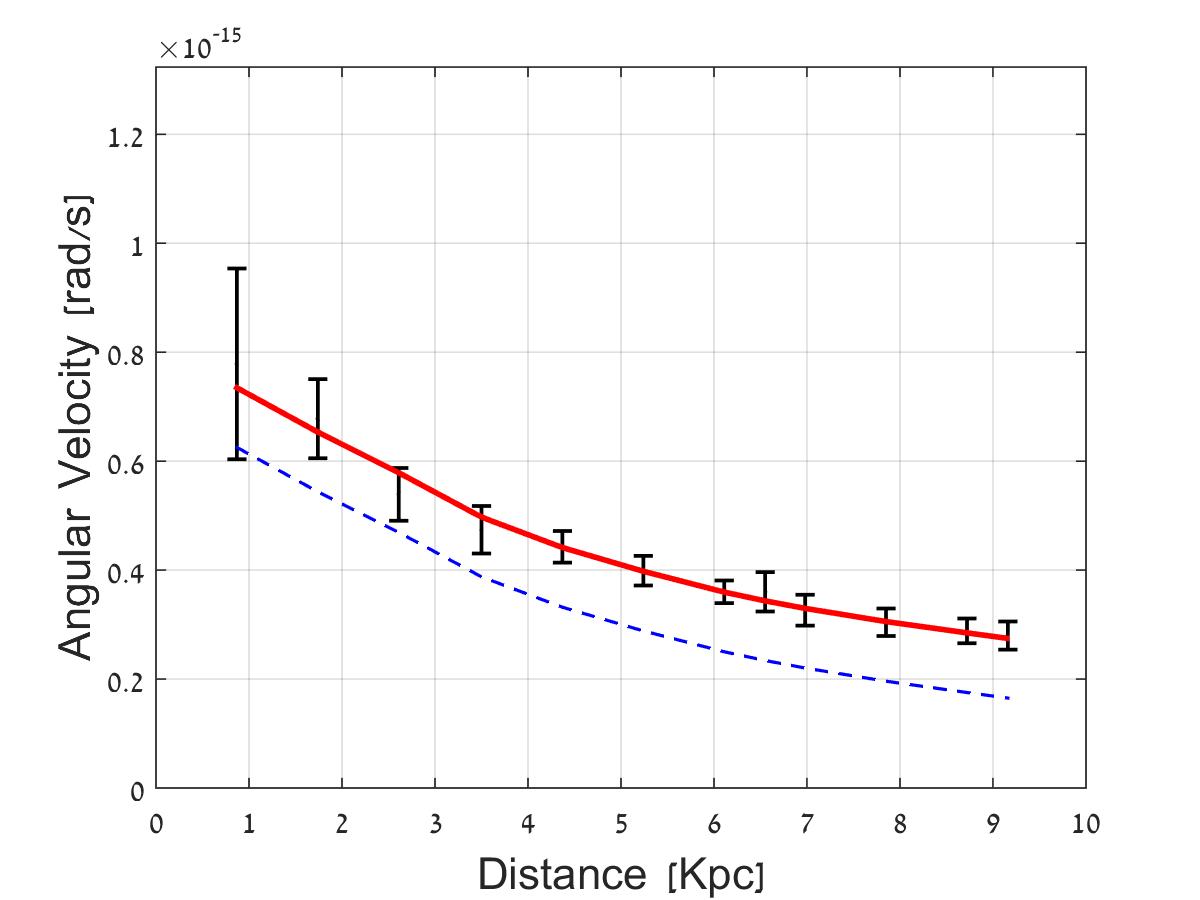}
\end{subfigure}
\begin{subfigure}{0.48\textwidth}
    \includegraphics[width=\textwidth]{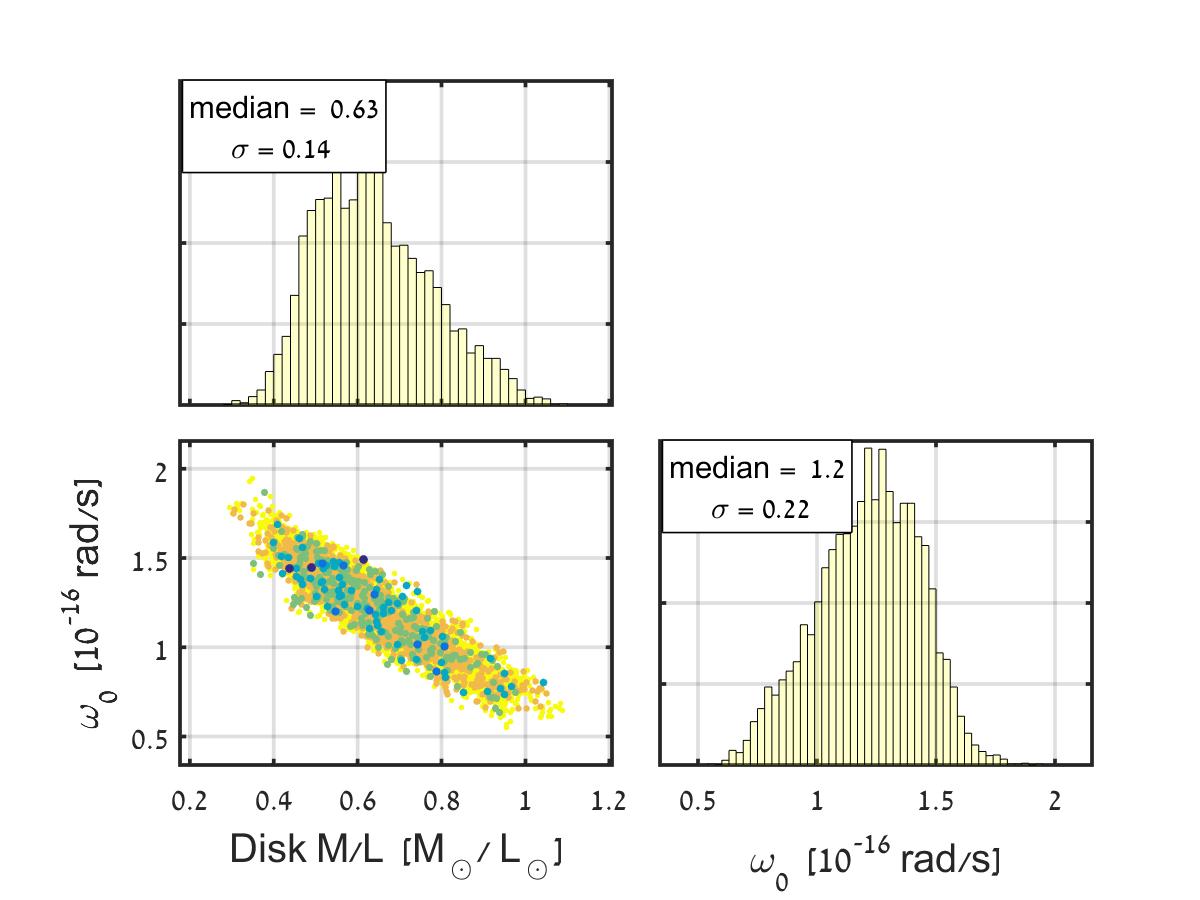}
\end{subfigure}
\caption{
Fitting the data of UGC-07089 with the new phenomenological model. The left panel plots the measured angular velocities (black points with error-bars) along with the model's prediction (red line). The baryonic contribution is displayed as a dashed blue line. The right panel presents a corner plot of the two model parameters. Bluer colours in the joint distribution indicate regions with higher likelihood. The complete set of AVC-fittings is available online (see Data-Availability Section). 
}
    \label{fig:MCMCexample}
\end{figure*}

Let us now dive into the analysis and explore the universality of the offsets. For this purpose we use the publicly-available SPARC database \citep{2016AJ_Lelli_152} together with the catalog of SPARC dark-halo models \citep{Li_mar_2020}. SPARC is a database that includes 175 nearby galaxies with new surface photometry at 3.6 $\mu$m and high-quality rotation curves. The publicly available catalog of \cite{Li_mar_2020} consists of 175 rotation-curve fits (for the entire SPARC sample) using seven different dark-halo profiles.

To confirm the universality of the offsets we introduce a phenomenological model that accounts for this property. The new model can be written as follows:
\begin{equation}
\omega_{\rm total}(r) = \omega_{\rm bar}(r) + \omega_{0}\label{eq:Property},
\end{equation}
where $\rm \omega_{total}$ is the angular velocity of an object orbiting a galactic center, $\rm \omega_{bar}$ is the angular velocity produced by the presence of a baryonic component, and $\rm \omega_{0}$ is a constant angular-velocity shift. Eq.\,\ref{eq:Property} is a data-driven relation. It is motivated by the potential presence of offsets in the data.

Our aim is to compare the new phenomenological relation (Eq.\,\ref{eq:Property}) with established approaches. Standard dark-halo profiles and the Modified Newtonian Dynamics model \citep[MOND,][]{1983ApJ_270_365M_MOND} already produce reliable fits to the data, so they may act as a baseline. If the new model (baryons + offset) could produce comparable results to these baseline models, it may be considered a viable approach. This would support the universality of angular-velocity offsets in disk galaxies.

The new model includes two free parameters: mass-to-light ($\rm M/L$) for the baryonic component and an additional constant $\rm \omega_{0}$. In order to quantify the goodness of fit of the new phenomenological model we perform a Markov Chain Monte Carlo (MCMC) simulation. Technical details are provided in the following paragraphs. For each galactic rotation curve, the MCMC estimates the best-fit model parameters, their errors and the corresponding $\rm \chi^{2}$.

\begin{figure*}
\centering
\begin{subfigure}{0.329\textwidth}
    \includegraphics[width=\textwidth]{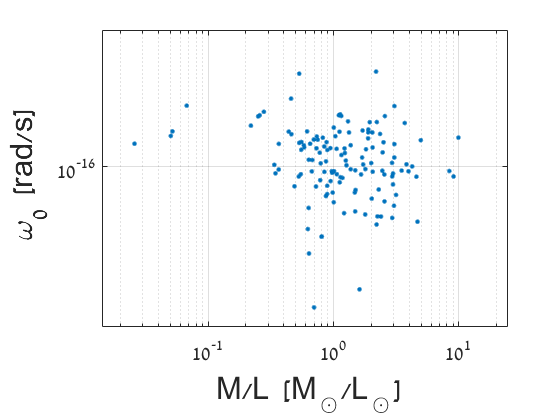}
\end{subfigure}
\begin{subfigure}{0.329\textwidth}
    \includegraphics[width=\textwidth]{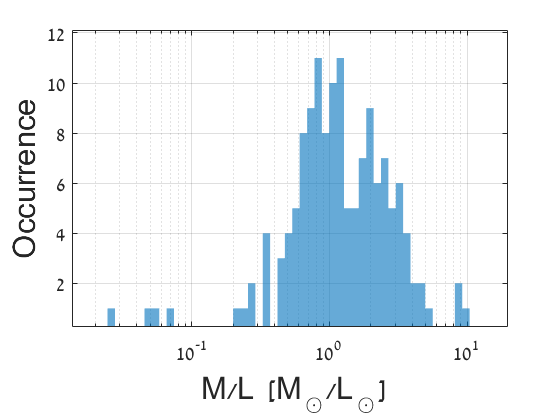}
\end{subfigure}
\begin{subfigure}{0.329\textwidth}
    \includegraphics[width=\textwidth]{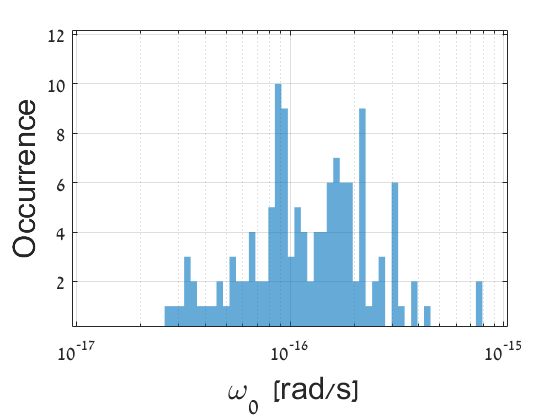}
\end{subfigure}
\caption{
A visual representation of the sample's best-fit model parameters. Left panel: A scatter plot of the two parameters. There is a weak but not significant anti-correlation between them ($\rm \rho_{pearson } = -0.1$, $p_{\rm value}=0.26$). Middle panel: Histogram of M/L values. The median value of log(M/L) is $\rm 0.06 \pm 0.04$. There are four outliers in the data with $\rm M/L < 0.1$, but within uncertainties, they are also consistent with higher values. Right Panel: Histogram of $\rm \omega_{0}$ values. The median value of $\rm log(\omega_{0})$ is $-15.97 \pm 0.51$. Eight points with $\rm \omega_{0} < 0$ are not shown, due to the logarithmic nature of the plot, but within uncertainties, they are also consistent with positive values.}
    \label{fig:Population}
\end{figure*}

Dark-halo models usually include three free parameters: $\rm M/L$ for the baryonic component and two more for the dark halo. In the current analysis we focus on two dark halo profiles: the cosmology-driven Navarro–Frenk–White \citep[NFW, ][]{1997ApJ_NFW} and the data-driven Burkert \citep{1995ApJ_447L_Burkert}. In order to quantify the goodness of fit of the dark-halo models we rely on  the catalog of \cite{Li_mar_2020}. For each galactic rotation curve, the catalog specifies the best-fit parameters and the corresponding $\chi^{2}$.

MOND includes one free parameter per galaxy: the $\rm M/L$ of the baryonic component. The critical acceleration $\rm a_0$, which is a universal constant in MOND theory, is treated as a global parameter. In order to quantify MOND's goodness of fit we also perform an MCMC simulation. The technical details are provided in Appendix~\ref{AppendixA}.

Since we prefer to use exactly the same data-sets when testing the competing models (i.e., NFW, Burkert, MOND and the current one), we convert the angular velocities of Eq. ~\ref{eq:Property} into actual velocities, as follows:
\begin{equation}
v(r) = v_{\rm bar}(r) + \omega_{0}r\label{eq:model}.
\end{equation}
As a result, all the different models are expressed in terms of rotational velocities, and are therefore applicable to the original rotation curve data.

The following paragraphs describe the new model and the MCMC analysis. First, to calculate the new model's baryonic contribution, we follow \cite{2016AJ_Lelli_152}. We write:
\begin{equation}
v_{\rm bar}(r)= \sqrt{(\rm M/L)\cdot v_{\rm disk}^{2}(r)+v_{\rm gas}^{2}(r)}\label{eq:baryon_model},
\end{equation}
where $\rm M/L$ is the stellar-disk mass-to-light ratio, $v_{\rm disk}(r)$ is the normalized stellar-disk contribution and $v_{\rm gas}(r)$ is the gaseous-disk contribution. The normalized stellar-disk and the gaseous-disk contributions are included in the SPARC database as data files. They were numerically derived from the corresponding light and neutral hydrogen distributions. More information on the process can be found in \cite{2002ARAA_40_Sanders}. 

Baryonic contributions are in fact included in all galactic rotation-curve models. In the general case, they may also contain a bulge term. However, in the current study, to avoid the complexity of adding an extra $\rm M/L$ parameter to the fittings (and the statistical analysis), or alternatively to avoid extra sensitivity to stellar population synthesis models, only galaxies with negligible bulge contributions were selected. This leaves us with 143 galaxies.

The next step is to apply priors to the parameters. For the $\rm M/L$ parameter we impose a prior of the form $\rm M/L$ > 0 to ensure that it maintains its physical meaning. For $\rm \omega_{0}$ we apply a non-informative uniform prior since no earlier knowledge is available. We then run the MCMC simulation through $10,000$ steps, using a Metropolis-Hastings algorithm and a burn-time of $800$ steps. To ensure the simulation's convergence, we apply a Gelman-Rubin test and make sure that the potential scale reduction factors are smaller than $1.2$ \citep{Hogg_2018, Gelman_Rubin_Test_1992}. The MCMC outputs are the estimated parameters and the corresponding $\rm \chi^{2}$ of each fit. An example of an MCMC fitting (converted to angular velocities for convenience) is given in Fig.~\ref{fig:MCMCexample}.


Looking at the figure, one may be interested in the particular influence of each parameter. Generally speaking, the role of the M/L parameter is to scale and modify the shape of the baryonic component, while $\rm \omega_{0}$ is responsible for the shift. Obviously, increasing M/L alone is not enough to fit the data, as the scaled shape will be severely distorted.

Another aspect would be the typical values of the parameters. Fig.\,\ref{fig:Population} displays the distributions of the parameters. Table\,\ref{table:Fitting_Results} provides the values, their errors, and additional statistical information. The median value of $\rm \omega_{0}$ is $\sim 10^{-16}$\,[rad/s], with a range of almost two orders of magnitude. The median value of $\rm M/L$ is $\sim \rm 1$ [$\rm M_{\odot} / L_{\odot}$], which is somewhat higher than expectations \citep{2014AJ_148_77M_McGaugh}. Appendix~\ref{AppendixB} tackles this concern by applying an additional fitting procedure, using different priors. Finally, there is a weak but not significant anti-correlation between the parameters ($\rm \rho_{pearson } = -0.1$, $p_{\rm value}=0.26$).

\begin{table}
\centering
\renewcommand{\arraystretch}{1.4}
\caption{Best-fit parameters and fitting information of Section \ref{sec:AVCs}}
\begin{tabular}{lccccc}
\hline
Galaxy & $\rm M/L$  & $\rm \omega_{0}$ & $\rm n$ & $\rm \chi^{2}$ & $\rm p_{value}$ \\
\hline
  &  [$\rm M_{\odot} / L_{\odot}$] & $\rm [10^{-16}\, rad /s]$  &   &   &   \\
\hline
NGC 3949 & $0.46^{+0.06}_{-0.04}$ & $2.03^{+0.46}_{-0.73}$ & 7  & 2.62  & 0.76 \\
NGC 3953 & $0.80^{+0.06}_{-0.05}$ & $0.22^{+0.20}_{-0.24}$ & 8  & 3.81  & 0.70 \\
NGC 3972 & $0.89^{+0.12}_{-0.11}$ & $1.44^{+0.30}_{-0.32}$ & 10 & 14.62 & 0.067 \\
\hline
\end{tabular}
\vspace{1mm}
\parbox{0.45\textwidth}{\small\textit{Note:} $\rm M/L$ denotes the stellar disk mass-to-light ratio, $\rm \omega_{0}$ is the angular-velocity offset, $n$ is the number of data points in the rotation curve, $\chi^{2}$ is the best-fit sum of normalized residuals, and $\rm p_{value}$ represents the corresponding probability. The full table is available online (see Data Availability Section).}
\label{table:Fitting_Results}
\end{table}

With both the MCMC and the catalog results in hand, the relative performance of the different models could be quantified. That is, the performance of the phenomenological model with respect to the established models. For this purpose we employ the Bayesian Information Criterion (BIC). BIC is a criterion for model selection that accounts for the trade-off between a model's fit and its complexity. It is suitable for comparing non-nested models when the data are assumed to be independent and normally distributed, with the noise being well characterized by measurement errors. The BIC score of a given model (with respect to a particular dataset) can be written as:
\begin{equation}
BIC = \chi^{2} + k \cdot ln(n)\label{eq:BIC},
\end{equation}
where $\chi^{2}$ is the sum of squares of the normalized residuals, $\rm k$ is the number of free model parameters and $\rm n$ is the number of data points.

We calculate the BIC difference (i.e. the $\rm \Delta BIC$ between NFW/Burkert/MOND and the phenomenological model) for the entire sample. A positive value indicates that the phenomenological model performs better than the baseline model and vice versa.

\begin{figure*}
\centering
\begin{subfigure}{0.329\textwidth}
    \includegraphics[width=\textwidth]{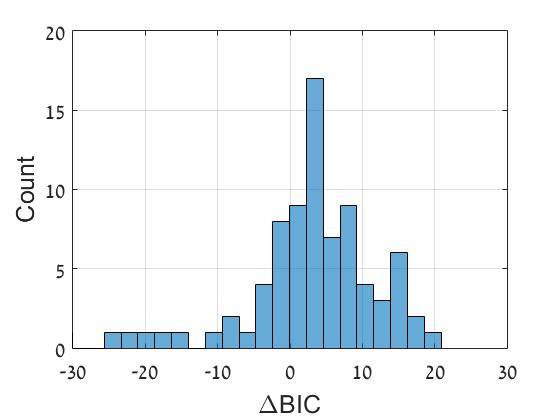}
\end{subfigure}
\begin{subfigure}{0.329\textwidth}
    \includegraphics[width=\textwidth]{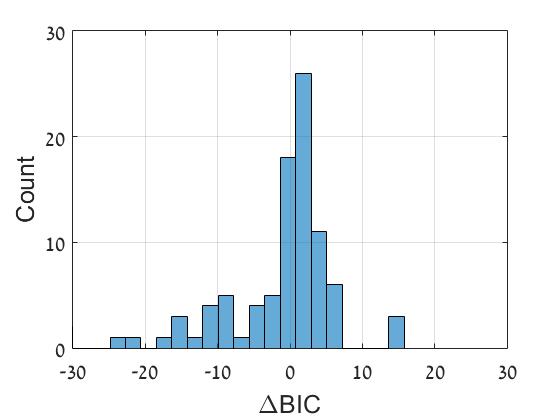}
\end{subfigure}
\begin{subfigure}{0.329\textwidth}
    \includegraphics[width=\textwidth]{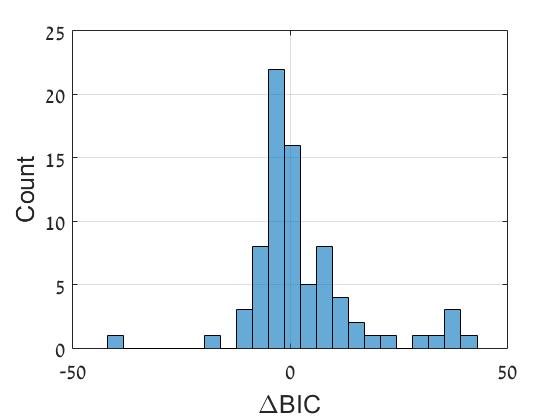}
\end{subfigure}
\caption{
Comparing the phenomenological model with established approaches. Left panel: A histogram of $\rm \Delta BIC$ values, where the comparison is relative to NFW. The median value of the histogram is 3.32 $\pm$ 0.75. The error on the median was calculated by Err = $\rm 1.48 * MAD / \sqrt N$, where N is the number of samples. Central panel: A comparison relative to the Burkert profile. The median value is 0.77 $\pm$ 0.36. Right panel: A comparison relative to MOND. The median value is -0.19 $\pm$ 0.68.
}
    \label{fig:Compare}
\end{figure*}

In the left panel of Fig.~\ref{fig:Compare} we compare the NFW model with the proposed phenomenological model. The sample's $\rm \Delta BIC$ results are plotted on a histogram. Generally speaking, the phenomenological model outperforms the NFW model. The vast majority of galaxies prefer the phenomenological model, with a median $\rm \Delta BIC$ of 3.32 $\pm$ 0.75. Such a value is unlikely to emerge from random fluctuations in the data as it is 4.4$\rm \sigma$ higher than zero.

Several rotation curves could not be properly fitted by the models (that is, fittings with $\rm \chi^{2}$ p-values smaller than 0.001). Since there is no point in comparing flawed results, we exclude from the BIC analysis 63 galaxies: 27 galaxies where the NFW fails, 13 where the phenomenological model fails, and 23 where both models fail.

On the central panel of Fig.~\ref{fig:Compare}, we plot a similar histogram, with the comparison being relative to Burkert's profile. In this case, the $\rm \Delta BIC$ test suggests that the models are equally credible, with slightly better results for the phenomenological model (median value of 0.77 $\pm$ 0.36). Flawed results were excluded as well (14 galaxies where Burkert fails, 20 where the phenomenological model fails, and 16 where both models fail).

In the right panel of Fig.~\ref{fig:Compare} we compare the current model to MOND. The $\rm \Delta BIC$ test suggests that the present model and MOND are equally credible, with a median value of -0.19 $\pm$ 0.68. It is necessary to note that there are several possible versions of MOND (e.g., \cite{2018_Li_Fitting_RAR_SPARC, 2021RAA_21_271W_MOND_SPARC}). Appendix~\ref{AppendixA} examines a few versions and their respective results. The adopted (main-text) result is based on \cite{2018_Li_Fitting_RAR_SPARC}. Flawed results were excluded as well (22 galaxies where MOND fails, 15 where the phenomenological model fails, and 31 where both models fail).

While the number of excluded galaxies appears to be high, the essential aspect here is that the numbers are comparable. In some cases the established models fail, whereas in others it is the phenomenological model that has difficulty fitting the data. The challenges in fitting the data depend on the accuracy of the data itself but also on the modeling of the baryonic component. These difficulties may include the extraction of surface photometry \citep{2016AJ_Lelli_152}, the bulge-disk decomposition \citep{2011ApJS_196_Simard} and the irregular morphologies of low surface-brightness galaxies.

Another way to look at the results is to divide the histograms into segments. It is generally agreed that $\rm \Delta BICs$ smaller than two are not significant. We therefore divide the histograms into three segments: $\rm \Delta BIC$ < -2, where the established model is preferred, -2 < $\rm \Delta BIC$ < 2, where both models are equivalent, and $\rm \Delta BIC$ > 2, where the phenomenological model is preferred. 

When compared to NFW, the new model is superior in 65\% of the cases, the models are equivalent in 15\% of the cases, and NFW is better in 20\% of the cases (73\%-27\% for a threshold of zero). When compared to Burkert, the new model is superior in 33\% of the cases, the models are equivalent in 37\% of the cases, and Burkert is preferred in 30\% of the cases (60\%-40\% for a zero threshold). Finally, when compared to MOND, the new model is superior in 36\% of the cases, the models are equivalent in 25\% of the cases, and MOND is preferred in 39\% of the cases (50\%-50\% for a zero threshold).

Ultimately, the current analysis suggests that the phenomenon of constant offsets in angular velocity curves is likely a universal property. The hypothesis is supported by the data as it appears comparable to standard dark-halo profiles and MOND. In Appendix \ref{AppendixE}, we present a further set of AVCs, spanning a wide range of galaxy sizes, in order to provide the reader with a broader perspective on the nature of AVCs.

\section{Deriving the Total Baryonic Mass}\label{sec:Masses}

It has been noted that angular velocity offsets may also prove useful in practical applications. In the current section, we would like to introduce such an application: a method that directly estimates the total baryonic masses of galaxies.

Baryonic masses are currently deduced from the data as part of a wider mass modeling (e.g. \cite{Li_mar_2020}). One typically selects a dark halo profile combined with an analytical (or numerical) term for the baryonic distribution. The model is then used to fit rotation curves and find the best-fit parameters. The total baryonic mass could be computed from the associated model parameters.

These methods are quite sensitive, as they rely on the chosen dark halo profile and the particular extraction of the baryonic distribution. In this section we aim to demonstrate that the mass could be computed in an independent manner. As long as angular-velocity offsets are taken as a universal property of disk galaxies, baryonic masses could be derived from the data without assuming how mass is distributed.

At a sufficiently large radius, i.e., in the outer parts of a galaxy, where the vast majority of baryonic mass resides well within this radius, the baryonic component should exhibit Keplerian behaviour. This notion raises the possibility to fit the outer parts of angular-velocity curves with a combined model: a Keplerian term and a constant term. The Keplerian term is an appropriate approximation to the baryonic component at large radii. The constant term represents the difference between the baryonic component and the observed data, as was shown in the previous section.

Let us derive the combined model. For a test particle located at the periphery of a galaxy, it may be possible to approximate the influence of the total baryonic mass ($M$) by using a point mass at the galactic center. Assuming a uniform circular motion, its acceleration towards the center is given by:
\begin{equation}
\omega^{2}r = GM / r^{2} \label{eq:KeplerianWDerive},
\end{equation}
where $\omega$ is the angular velocity induced by the baryonic component, $\rm M$ is the baryonic mass, $\rm G$ is the gravitational constant and $\rm r$ is the radius of the circular motion. Extracting $\rm \omega$ from Eq.~\ref{eq:KeplerianWDerive} and accounting for the constant offset, one obtains:
\begin{equation}
\omega_{total}(r) = \sqrt{\rm GM} / r^{1.5} + \omega_{0} \label{eq:OuterModel},
\end{equation}
where $\rm \omega_{0}$ and $\rm M$ are the free parameters of the model. Note that this model does not rely on measured quantities (unlike Eq.~\ref{eq:baryon_model} for example) and is therefore insensitive to the difficulties associated with their calculation.

This straightforward analytical model could be applied to observed data from the outskirts of galaxies (i.e. the outer parts of AVCs). For this purpose we employ (once again) the SPARC sample \citep{2016AJ_Lelli_152}. The outer data of each galaxy in the sample are extracted and used by the model described in Eq.~\ref{eq:OuterModel}.

It is essential to define what is considered "far enough". That is, to define where the Keplerian approximation is adequate. Our approach involves two constraints. The first relies on the effective radius ($\rm R_{eff}$), which encompasses half of the total luminosity. We consider data points with a radius larger than $\rm 2R_{eff}$. This means that the vast majority of light (and mass) is well within this radius. In order to avoid the sensitivity of relying on a single quantity, we impose a second restriction. It is based on the maximal radius ($\rm R_{max}$) which represents the last and farthest point of the rotation curve. We consider only data points with a radius larger than $\rm 0.5 R_{max}$. Finally, we consider AVCs with at least three data points after the above filtering. As a result, 128 AVCs remain.

\begin{figure}
		\includegraphics[width=\columnwidth]{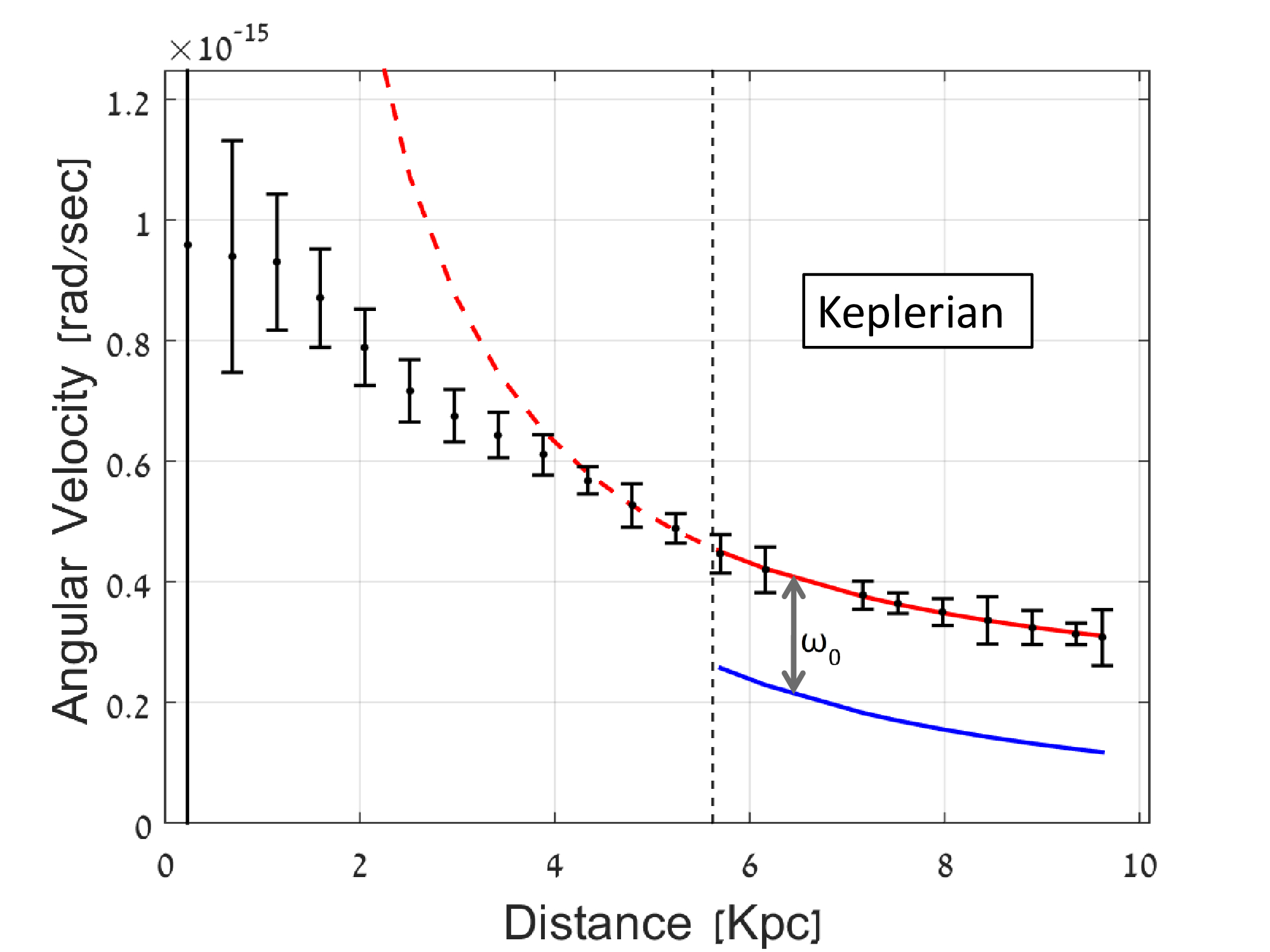}
    \caption{The observations (black error bars), the Keplerien baryonic contribution (blue line) and the corresponding model (red curve) of NGC-100. Note that the relevant part of the model is given in solid red, while the inner part (dashed red) is presented for clarity. The Keplerian cutoff (as defined in the main text) is indicated by a dashed black line.}
    \label{fig:OuterAVCs}
\end{figure}

Fig.~\ref{fig:OuterAVCs} presents an example of an "outer-AVC" fitting. The model is indicated by a solid red line. The dashed red line, displayed for clarity, represents the inner part of the model, i.e. the region where the Keplerien assumptions are not valid. Ultimately, the fitting procedure determines the values of $\rm \omega_{0}$ and $\rm M$ for each galaxy in the sample. The values of $\omega_{0}$ are not precisely identical to those obtained in Section\,\ref{sec:AVCs} since the fitting approach is different in this case.

Finally, the derived baryonic masses are given in Table \ref{table:Outer_Fitting_Results}. The median value of the population is $\rm 1.66 \pm 1.1 \ [10^{10}\ M_{\odot}]$, with the masses spanning $\rm 4$ orders of magnitude. The full table is available online (see Data Availability).

\section{A Relation between Galaxies and their Host Dark Halos} \label{sec:Relation}

Dark-halo profiles typically provide reliable fits to rotation curves. In some cases, however, a degeneracy between their two (or three) parameters appears during the fitting process \citep{1985ApJ_295_van_Albada, 2005ApJ_619_Dutton, MNRAS_2006_Fardal}. As was demonstrated in Section \ref{sec:AVCs}, the phenomenological model can also provide credible fits to RCs (or AVCs). Since the phenomenological model includes only one additional parameter (in addition to the $\rm M/L$), it provides an opportunity to design a dark halo profile with a single parameter.

In order to design such a profile, let us begin with the dynamics. Let us examine the influence of the different mass components on a test particle. The inward force acting on a body (which is located on the galactic plane at a distance $\rm r$ from the center) could be divided into its baryonic and dark contributions:
\begin{equation}
F(r) = F_{\rm bar}(r) + F_{\rm dark}(r)\label{eq:Forces},
\end{equation}
where $\rm F(r)$ is the total inward force acting on the body, while $\rm F_{\rm bar}(r)$ and $\rm F_{\rm dark}(r)$ are the inward forces created by the baryonic and dark components alone.

Since $\rm F \propto v^2$ (in uniform circular motions at a given radius), Eq.~\ref{eq:Forces} can also be given as:
\begin{equation}
v_{\rm total}^{2}(r) = v_{\rm bar}^{2}(r) + v_{\rm dark}^{2}(r)\label{eq:sVelocities},
\end{equation}
where $\rm v_{total}$ is the total circular velocity of an object orbiting a galactic center, while $\rm v_{bar}$ and $\rm v_{dark}$ are the circular velocity components due to the baryonic and dark components alone. Eq.~\ref{eq:sVelocities} is widely used in rotation curve fitting. It does not assume how the baryonic or the dark halo components are distributed (e.g., spherical or flat) but only requires that the net force of each component is in the inward direction.

\begin{table}
\centering
\renewcommand{\arraystretch}{1.4}
\caption{Best-fit parameters and fitting information of Section \ref{sec:Masses}.}
\begin{tabular}{lccccc}
\hline
Galaxy & $\rm M$ & $\rm \omega_{0}$ & $\rm n$ & $\rm \chi^{2}$ & $\rm p_{value}$ \\
\hline
  & [$\rm 10^{10}\ M_{\odot}$] & $\rm [10^{-16}\, rad /s]$  &   &   &   \\
\hline
NGC 3949 & $1.23^{+0.59}_{-0.46}$ & $3.77^{+1.14}_{-1.20}$ & 3  & 0.0087  & 0.92 \\
NGC 4100 & $7.59^{+0.33}_{-0.38}$ & $0.63^{+0.08}_{-0.06}$ & 13  & 6.64  & 0.83 \\
NGC 4214 & $0.34^{+0.14}_{-0.12}$ & $1.58^{+0.93}_{-0.91}$ & 7 & 0.55 & 0.99 \\
\hline
\end{tabular}
\vspace{1mm}
\parbox{0.45\textwidth}{\small\textit{Note:} $\rm M$ denotes the total baryonic mass, $\rm \omega_{0}$ is the angular-velocity offset, $n$ is the number of data points in the outer part of the rotation curve, $\chi^{2}$ is the best-fit sum of normalized residuals, and $\rm p_{value}$ represents the corresponding probability. The full table is available online (see Data Availability Section).}
\label{table:Outer_Fitting_Results}
\end{table}

A similar argument can be made when angular-velocity curves are employed. Since $\rm F \propto \omega^2$ (in uniform circular motions at a given radius) one may similarly write:
\begin{equation}
\omega_{\rm total}^{2}(r) =  \omega_{\rm bar}^{2}(r) +  \omega_{\rm dark}^{2}(r)\label{eq:sAngVelocities},
\end{equation}
where $\rm \omega_{total}$ is the total angular velocity of the orbiting object, while $\rm \omega_{bar}$ and $\rm \omega_{dark}$ are the angular velocity components due to the baryonic and dark components alone.

The reader is reminded that Eq.~\ref{eq:Property} also describes total angular velocities. It was suggested that angular-velocity curves could be separated into a baryonic component, $\rm \omega_{bar}(r)$, and an additional offset $\rm \omega_{0}$. By inserting Eq.~\ref{eq:Property} into Eq.~\ref{eq:sAngVelocities} one obtains:
\begin{equation}
\omega_{\rm dark}^{2}(r) =  \omega_{0}^{2} +  2\omega_{0}\omega_{\rm bar}(r) \label{eq:Universalrelation}.
\end{equation}

Eq.~\ref{eq:Universalrelation} connects the angular velocities induced by a dark halo and those induced by the corresponding baryonic component. Note that when the baryonic model is available (i.e. with a particular $\rm M/L$ value), the dark distribution is characterized by a single parameter.

Let us take a step further and explicitly present the halo's mass-density distribution. That is, a halo distribution that induces a gravitational field (and in turn, velocity fields) that are compatible with Eq.~\ref{eq:Universalrelation}. The complete derivation is provided in Appendix \ref{AppendixC}. The derivation assumes a spherically symmetric halo, and the derived density profile is given by:
\begin{equation}
\rho_{\rm dark}(r) = \frac{\omega_{0}^{2}}{4\pi G}(3 + 4\frac{v_{\rm bar}(r)}{\omega_{0}r} + 2\frac{v'_{\rm bar}(r)}{\omega_{0}}) \label{eq:Rho_dark},
\end{equation}
where $\rm \omega_{0}$ is the constant offset, $\rm v_{\rm bar}(r)$ is the baryonic model for the rotational velocities, and $\rm v'_{\rm bar}(r)$ is the first derivative of $\rm v_{\rm bar}(r)$ with respect to r.

Eq.~\ref{eq:Rho_dark} describes a new dark-halo profile. The halo is characterized by one free parameter, $\rm \omega_{0}$, which represents a constant offset in the data. The free parameter may be replaced with a characteristic density of the form $\rm \rho_{0} = \omega_{0}^{2} / 4\pi G$. For a typical value of $\rm \omega_{0}\sim10^{-16}$\,rad/s one would get a typical value of $\rm \rho_{0}\sim 10^{-23}$\,kg/m$^{3}$. This is $\sim4$ orders of magnitude higher than the average density of dark matter in the universe \citep{PhysRevLett86_385_baryon_density, 2020AA_641A_6P_Planck}, which is expected for a typical value in dense regions.

Furthermore, the halo structure is directly connected to the baryonic velocity distributions. This may yield insightful results. For example, if the inner region of the baryonic distribution could be approximated by $\rm v_{bar} \propto r$ then a cored dark halo would appear (i.e. a halo with a constant central density). On the other hand, if the inner region could be associated with a velocity term of the form $\rm v_{bar} \propto \sqrt r$ then the inner density profile would be approximated by $\rm \rho_{dark} \propto 1 / \sqrt r$.

\begin{figure}\includegraphics[width=\columnwidth]{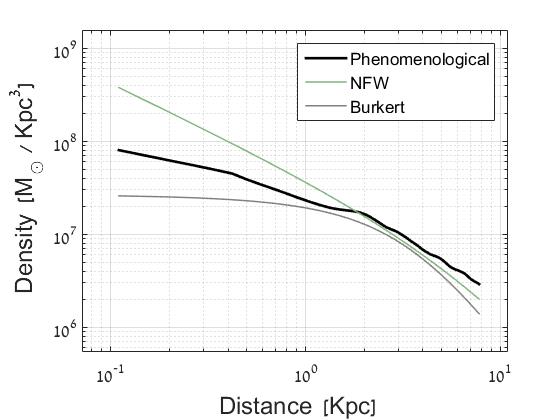}
\caption{A comparison of dark-halo profiles. The data-driven (cored) Burkert, the cosmology-driven (cuspy) NFW, and the current phenomenological profile are plotted together. The phenomenological profile is intrinsically related to the baryons. In this example, the profiles' parameters (or intrinsic terms) are tuned to NGC-7793.}
    \label{fig:DM_Halos}
\end{figure}

This kind of behaviour may further be explored either by using analytical models or by applying numerical models of particular galaxies. In Fig.~\ref{fig:DM_Halos} we employ the second approach and display the new profile's density distribution for a particular galaxy (NGC-7793). The value of $\rm \omega_{0}$  (and the numerically-derived baryonic velocities) are the same as in Section \ref{sec:AVCs}. The profile is plotted alongside the associated NFW and Burkert profiles, with their parameters taken from \cite{Li_mar_2020}. Interestingly, the inner slope of the new density profile lies between Burkert and NFW with a value of $\rm -0.7$. The surrounding regions roughly appear to resemble the established profiles.

The coupling between dark and baryonic distributions in spiral galaxies has been extensively discussed in the literature \citep{1996MNRAS_281_Persic, 2000ApJ_537L_Salucci, 2009AA_493_Swaters, 2014Galax_2_McGaugh}. The current study introduces an explicit relation between these two distributions.

\section{Summary and Discussion}

The present study introduces a new class of rotation curves: the angular-velocity curves. At first sight, it seemed arbitrary to introduce such a category since no additional information on the galaxies was offered. However, switching to angular-velocities proved useful due to the intrinsic property of constant offsets.

It has been demonstrated that a typical angular-velocity curve could be decomposed into a baryonic contribution and a constant shift. In other words, the observed curve and its baryonic-based prediction typically show similar patterns when the angular-velocity view is adopted. Throughout this work, this property was referred to as the \textit{constant offsets} property.

The generality of the property was analysed by comparing it to well-established models. Our analysis seem to support the presence of offsets in disk galaxies, as the property performs equally well (or better) than the standard models. Compared to a Burkert profile, it is preferred in 60\% of the cases, relative to NFW it is superior in 73\% of the cases, and relative to MOND it is equally credible. The analysis was carried out by applying the SPARC sample, which consists of 175 rotation curves. As a potential next step, it would be beneficial to test the hypotheses on larger samples, like MaNGA \citep{2017AJ_154_86W_Wake}.

In adopting the current property, a concern may emerge. A constant angular velocity term is equivalent to a circular velocity that increases linearly with radius. At a sufficiently large distance, the linear term may become dominant and lead to increasing rotation curves in the outer parts of galaxies. Yet, most RCs are observed to flatten out in these regions \citep{2001ARAA_39_137S}.

The resolution of this concern is fairly straightforward. A visual inspection of the full sample reveals that in most RCs the rate at which the baryonic term is declining is approximately the same as the rate at which the linear term rises over very long distances. This behaviour produces flat, slightly rising, or slightly falling curves at the outer parts of galaxies. In Appendix \ref{AppendixD} we present this concern and its resolution in more detail, as well as a few examples.

A beneficial side-effect of the offsets is the introduction of a method that estimates galaxies' total baryonic masses. By focusing on the peripheral regions, where the baryonic component exhibits a Keplerian behaviour, it was possible to design an analytical model for the rotational velocities. The proposed model for the velocities does not assume how the mass is distributed. The parameters that define the distribution of the dark (or the baryonic) profiles are not necessary. The only model-parameters are the total baryonic mass and the offset. This enables the calculation of baryonic masses while not being affected by baryonic modeling. Potentially, this method could be applied in the future to a large number of galaxies to estimate their baryonic masses. 

Other routes for investigation, in this context, may include the compatibility of the masses with a baryonic Tully-Fisher relation \citep{2000ApJ_533L_McGaugh}. Since the outer flat parts of rotation curves can be well-fitted with the current model (see examples in Appendix \ref{AppendixD}), it may be interesting to correlate the baryonic masses with their corresponding typical flat velocities, assuming the latter represent the Keplerian region. Another direction would be the sensitivity of the current method to the degeneracy between the disk and the halo \citep{2005ApJ_619_Dutton}. \\

The realization that baryonic and observed angular velocities exhibit similar behaviour may also lead to a curious interpretation of the phenomenon. Even though the baryonic predictions consistently underestimate the data, it seems that the baryons are somehow "aware" of the overall dynamics. Such a similarity may suggest that the dynamics are indeed described by the baryons, while there is some additional underlying mechanism responsible for the offsets.

A possible mechanism of such a kind may be linked to a rotation of the reference frame \citep{2021Galax_GZ}. More accurately, a rotation of the local inertial frame where the model is applicable, relative to the frame where the observations are given. Such a rotation would induce an additional constant angular-velocity term in the data.

Technically, there is no inherent prohibition against such a setting of the inertial frame. However, such a view would require a shift from the $\rm \Lambda CDM$ cosmology since it does not involve dark halos. The data, according to this mechanism, can be fully explained by the baryonic component and the dragging of the inertial frame. In that sense, the proposed interpretation is radical. One of the main challenges of such a view would be to explain how, in a systematic manner, the local inertial frame of every galaxy is "dragged" relative to its observational frame.

A more conservative strategy would be to design the particular form of a dark halo distribution that induces constant offsets in the data. By designing such a profile, starting from the dynamical equations, it turned out that the halo is directly related to the baryonic distribution. That is, the new dark-halo profile is also a phenomenological (data-driven) dark-baryonic relation. Given a model for the baryonic distribution (e.g., through photometry-based methods), one may directly extract the dark-halo mass distribution of a particular galaxy (Eq.~\ref{eq:Rho_dark}).

The coupling between baryonic and dark matter in galaxies is a fundamental aspect of cosmic structure formation \citep{Numerical_simulation_2023_AA_dark_halos, 2017MNRAS_466_Baryonic_dark_relations}. Observational studies, particularly those analysing galaxy rotation curves, have revealed that the distribution of baryonic matter can significantly influence the morphology of dark matter halos. For instance, high surface brightness galaxies exhibit more contracted and cuspy dark matter profiles, suggesting a strong coupling between baryonic distribution and halo shape \citep{Coupling_DM_Baryons_2022_AA}. Similarly, analyses employing Einasto halo fits have revealed correlations between the core properties of dark matter halos and the extent of the baryonic component \citep{2019AA_623A_123G}. Since the current study introduces an explicit relationship between the two distributions, several observational studies in small-scale cosmology (e.g., \cite{PhysRevD_110_063028_Benisty}) may benefit from using it directly.

The emergence of an explicit relation may require further exploration. One future possibility is to analyse the relation by comparing it to standard dark halo profiles. Since each profile has its own distinctive characteristics, it may be illuminating to visualize the different profiles together as in Fig.~\ref{fig:DM_Halos}. This may be particularly relevant from the perspective of the core-cusp problem \citep{2010AdAst_5D_de_Blok}. Let us clarify that a deeper understanding of Eq.~\ref{eq:Rho_dark} is equivalent to understanding the offsets.

Another subject to consider is the connection between the current phenomenology and MOND. Although MOND is based on theoretical assumptions, its application to RCs may be considered semi-empirical. This is due to the flexibility afforded by the selection of the unknown function $\mu (a / a_{0})$. In that sense, both MOND and the current model are semi-empirical prescriptions for rotation-curve fitting based on the corresponding baryonic component. In fact, both methods are highly sensitive to baryonic modeling. The current model uses a constant offset, which preserves any features in the baryonic term, while MOND uses only the baryonic term, together with the chosen $\mu(a / a_{0})$ \citep{1983ApJ_270_365M_MOND}.

There are also notable differences between the two models. First, for a given MOND function $\mu$ and a global acceleration value $a_0$, MOND requires only one free parameter per galaxy, that is, $\rm M/L$. The current model requires both $\rm M/L$ and an offset for each galaxy (see Appendix \ref{AppendixA}). The second difference is more basic. Because MOND is a theory of gravity, acceleration is the most crucial quantity. The present phenomenology, however, suggests that angular velocities are the important metric to investigate.

Finally, we hope that the newly-discovered phenomenological pattern will be further explored and analysed. It could prove valuable in additional applications, while the relation between dark and baryonic distributions may contribute to our understanding of disk galaxies.

\section*{Acknowledgements}
The authors wish to thank David Benisty for his helpful suggestions and the anonymous referee for his thoughtful questions and valuable feedback. This research was supported in part by the Israel Science Foundation (ISF), grant No. 1404/22.

\section*{Data availability}
The rotation-curve data underlying this study are available in the DATA SECTION, at \url{http://astroweb.cwru.edu/SPARC/}. It includes extended 21-cm rotation curves and B-band photometry for the sample of galaxies analysed in this work. The complete set of figures supporting the findings of this study is openly available at \url{https://doi.org/10.5281/zenodo.17392323}. These include figures and tabulated data of the model's fitting results.


\bibliographystyle{mnras}
\bibliography{Bib/MyBib}


\appendix

\section{A comparison with MOND}\label{AppendixA}

One of the main goals of the present study is to establish the universality of constant offsets. In Section \ref{sec:AVCs} we compare the current phenomenological model (i.e., of constant offsets) with existing models. A comparison with MOND, one of the earliest and most established models, may require several implementations. Therefore, the current appendix describes several different MOND versions and their corresponding results.

First, we follow \cite{2021RAA_21_271W_MOND_SPARC} and choose a simple interpolation function of the form:
\begin{equation}
\rm \mu(x) = \frac{x}{1+x} \label{eq:MOND_Rel_1},
\end{equation}
where $\mu$ is the MOND interpolation function, and $\rm x = a/a_{0}$, is the ratio between actual accelerations and the critical acceleration. For the critical acceleration we adopt a value of $\rm a_{0} = 1.2 \cdot 10^{-10} [m /s^{2}]$ \citep{1991MNRAS_249_523B, 2018_Li_Fitting_RAR_SPARC}.

To calculate MOND's rotational velocities (i.e., to produce a MOND model), one needs a conversion formula translating from Newtonian velocities to MOND. The current formula can be directly derived from Eq. \ref{eq:MOND_Rel_1} and is explicitly given in \cite{2021RAA_21_271W_MOND_SPARC}.

Next, with a MOND model in hand, we apply an MCMC simulation to fit the model to the data. We use a single parameter (the stellar disk $\rm M/L$) with a positive prior ($\rm M/L > 0$) to find the best-fit value for each galaxy and its corresponding $\chi^{2}$. The rest of the MCMC details are the same as in Section \ref{sec:AVCs}.

\begin{figure*}
\centering
\begin{subfigure}{0.329\textwidth}
    \includegraphics[width=\textwidth]{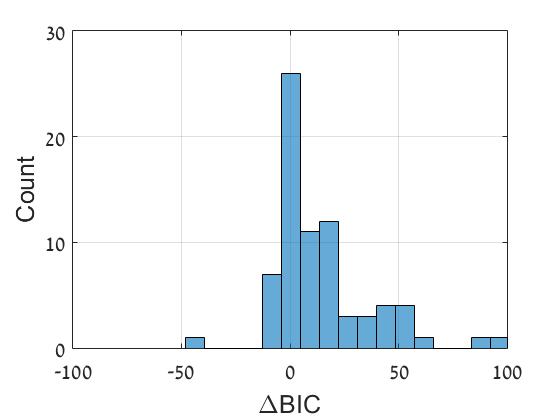}
\end{subfigure}
\begin{subfigure}{0.329\textwidth}
    \includegraphics[width=\textwidth]{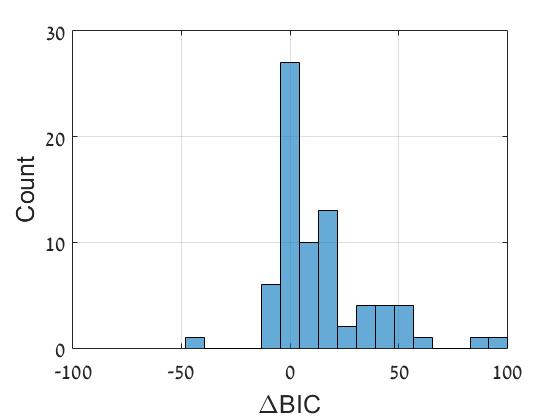}
\end{subfigure}
\begin{subfigure}{0.329\textwidth}
    \includegraphics[width=\textwidth]{Figures/Comparisons/GZvsMOND_RAR_DI.jpg}
\end{subfigure}
\caption{
Comparing the phenomenological model with MOND. Left panel: A histogram of $\rm \Delta BIC$ values, where the comparison is relative to a simple MOND (see the main text). The median value of the histogram is 8.31 $\pm$ 1.74. Central panel: A comparison relative to a RAR-MOND profile. The median value is 8.26 $\pm$ 1.73. Right panel: A comparison relative to RAR-MOND with adjustable distance and inclination. The median value is -0.19 $\pm$ 0.68.
}
    \label{fig:CompareMOND}
\end{figure*}

The MOND MCMC results are then used for a statistical comparison. We follow the scheme described in Section \ref{sec:AVCs} to calculate the $\rm \Delta BICs$ between MOND and the current model. The results are shown in the left panel of Fig.~\ref{fig:CompareMOND}. The $\rm \Delta BIC$ test already accounts for the fact that MOND requires only one free parameter.

According to the test, the present phenomenological model is preferred over MOND. The median value is 8.31 $\pm$ 1.74 with $65\%$ of the galaxies produce $\rm \Delta BIC > 2$. Flawed results were excluded from the test (43 galaxies where MOND fails, 6 where the phenomenological model fails, and 30 where both models fail).

The next step is applying a different interpolation function. We rely on \cite{2018_Li_Fitting_RAR_SPARC} and apply the following formula to convert the accelerations:
\begin{equation}
\rm a = \frac{a_{bar}}{1-e^{-\sqrt{\frac{a_{bar}}{a_{0}}}}} \label{eq:MOND_Rel_2},
\end{equation}
where $\rm a$ is the MOND acceleration and $\rm a_{bar}$ is the baryonic Newtonian acceleration. The conversion in Eq. \ref{eq:MOND_Rel_2} is based on the radial-acceleration relation (RAR).

We then apply precisely the same scheme (MCMC, statistical analysis) to compare MOND and the current model. The results are shown in the central panel of Fig.~\ref{fig:CompareMOND}. The histogram is almost unaffected by the two different MOND models. In fact, the vast majority of MOND fits are almost indistinguishable when visually inspected. In this case, the present phenomenological model is preferred over MOND, with a median value of 8.26 $\pm$ 1.73 and with $65\%$ of the galaxies producing $\rm \Delta BIC > 2$. The excluded galaxies are similar to the previous test.

The last step is to treat the distance and inclination as free parameters. Following \cite{2018_Li_Fitting_RAR_SPARC} we allow the distance and inclination to be adjusted while priors are set according to their known uncertainties. We also use a prior for the $\rm M/L$ parameter, in accordance with stellar population synthesis models and precisely as in \cite{2018_Li_Fitting_RAR_SPARC}. The extra flexibility considerably improves MOND results.

To enable a proper comparison, it is necessary to apply these adjustments to the phenomenological model as well. The details of the later analysis (that is, of the adjusted phenomenological model) are presented in Appendix \ref{AppendixB}. The results of the comparison are shown in the right panel of Fig.~\ref{fig:CompareMOND}.

According to the last test, the present phenomenological model and MOND are equally credible. The median value is -0.19 $\pm$ 0.68, with $36\%$ of the galaxies produce $\rm \Delta BIC > 2$ and $39\%$ produce $\rm \Delta BIC < -2$. Throughout the main text, this final test is presented. Flawed results were excluded from the test (22 galaxies where MOND fails, 15 where the phenomenological model fails, and 31 where both models fail).

\section{An additional Fitting Method}\label{AppendixB}

The current appendix offers an additional method to fit the phenomenological model to the data. Following previous studies, and especially \cite{2018_Li_Fitting_RAR_SPARC}, the distance and inclination of a galaxy may be treated as free parameters, with priors assigned according to their uncertainties. This affords some extra flexibility in the fitting process and reflects an inherent uncertainty in the modeling.

In our case, it would be worthwhile to investigate whether such flexibility could produce lower $\rm M/L$ values. As discussed in Section \ref{sec:AVCs}, the current $\rm M/L$ values, obtained from the baseline fitting, are somewhat higher than expected by stellar population synthesis models \citep{Bell_2001, 2012ApJ_744_17M_ML}. Another motivation to implement the current method is to enable the third statistical test in Appendix \ref{AppendixA}.

In order to implement the outlined method we follow \cite{2018_Li_Fitting_RAR_SPARC} and adjust the baryonic component through:
\begin{equation}
\rm v_{bar}^{'} = \sqrt\frac{D}{D_{0}} \ {v_{bar}} \label{eq:D_D0},
\end{equation}
where $\rm D$ is the adjustable distance and $\rm D_{0}$ is the default value of the distance as given in SPARC data tables. The added parameter $\rm D$ is not entirely flexible. It is constrained by a Gaussian prior with its standard deviation being equal to the specified distance uncertainty.

A second additional parameter is disk inclination \citep{2018_Li_Fitting_RAR_SPARC}. When the disk inclination is adjusted, the observed velocities are transformed by:
\begin{equation}
\rm v_{obs}^{'} = \frac{sin(i_{0})}{sin(i)} \ {v_{obs}} \label{eq:i_i0},
\end{equation}
where $\rm i$ is the adjustable inclination and $\rm i_{0}$ is the default value of the inclination as given in SPARC data tables. The inclination is also constrained by a Gaussian prior with its standard deviation equal to the specified inclination uncertainty.

For the $\rm M/L$ parameter we use a log-normal prior with a mean value of $\rm 0.5 \ [M_{\odot}/L_{\odot}]$ and a standard deviation of $\rm 0.15$ dex, the latter being a slightly weaker than the commonly used $0.1$-dex prior \citep{2012ApJ_744_17M_ML, Li_mar_2020}. We then run the MCMC simulation (as specified in Section \ref{sec:AVCs}) to obtain the best-fit parameters and their corresponding $\chi^{2}$ values.

The fitting results are quite informative. The number of unsuccessful fittings (with p-value < 0.001) increased from 36 to 46. It reflects galaxies where the $\chi^{2}$ is not significantly improved by the extra flexibility, but the higher number of parameters reduces their p-value. A reduction in p-value may also occur in galaxies with excellent fittings, where extra flexibility is not required. On the other hand, a few fittings (e.g., UGC-00731) have been considerably improved, with inclination being the dominant parameter.

\begin{figure}
	\includegraphics[width=\columnwidth]{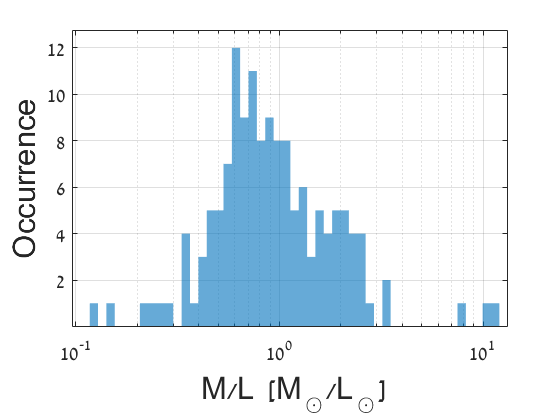}
    \caption{Histogram of updated $\rm M/L$ values. The median value of log($\rm M/L$) is  $\rm -0.17\pm 0.05$.} 
    \label{fig:ML_append}
\end{figure}

Fig.~\ref{fig:ML_append} draws an updated histogram of $\rm M/L$ values. The median value of log($\rm M/L$) is -0.17 with a standard error of 0.05. This is substantially lower than the values in Fig.~\ref{fig:Population}, but still a bit higher than expectations \citep{2012ApJ_744_17M_ML}. Whether this reflects a genuine property of the $\rm M/L$ values or a weakness in the phenomenological model deserves further exploration.

A final notable point deals with inclination values. Several galaxies exhibit phenomenological-model inclinations that strongly disagree with those obtained from a similar MOND analysis (i.e. those in \cite{2018_Li_Fitting_RAR_SPARC}). As an example, UGC-08837 is well fitted by both models, with the phenomenological model producing $\rm i/i_{0} = 1$ and MOND predicting $\rm i/i_{0} = 0.73$. Future examination of the observed inclinations may provide insight into this conflict.

\section{The Dark-Halo Density Profile}\label{AppendixC}

The following appendix derives the halo's density distribution presented in the main text (Eq. \ref{eq:Rho_dark}). Given a uniform circular motion, the acceleration of a test particle towards the center of a galaxy, in the presence of a spherically-symmetric halo, is given by:
\begin{equation}
\omega_{\rm dark}^{2} \cdot r = \frac{GM_{\rm dark}(r)}{r^{2}}  \label{eq:DarkAttr1},
\end{equation}
where $\rm \omega_{dark}$ is the angular velocity component due to the dark halo, $\rm r$ is the radius, and $\rm M_{dark}(r)$ is the halo's enclosed mass. Extracting the halo's mass from Eq. \ref{eq:DarkAttr1} gives:
\begin{equation}
M_{\rm dark}(r) = \frac{\omega_{\rm dark}^{2}(r) \cdot r^{3}}{G} \label{eq:DarkAttr2}.
\end{equation}

Next, by adding the phenomenological relation \ref{eq:Universalrelation} into Eq. \ref{eq:DarkAttr2}, one may connect the halo's mass distribution with the baryonic angular-velocities as well. This results in:
\begin{equation}
M_{\rm dark}(r) = \frac{(\omega_{0}^{2} + 2\omega_{0} \omega_{\rm bar}(r) ) \cdot r^{3}}{G} \label{eq:DarkAttr3},
\end{equation}
where $\rm \omega_{0}$ is the constant offset and $\rm \omega_{bar}(r)$ is the baryonic angular-velocity distribution.

The next step is to derive the density profile. Assuming a spherically-symmetric halo, one may write:
\begin{equation}
\rm dM = \rho(r) \cdot 4\pi r^2  dr \label{eq:Mass_Dens},
\end{equation}
where $\rm dM$ is a differential mass element of the halo and $\rm \rho(r)$ is the halo's density distribution. An equivalent formulation would be:
\begin{equation}
\rm M'(r) = \rho(r) \cdot 4\pi r^2 \label{eq:Mass_Dens2},
\end{equation}
where $\rm M'(r)$ is the derivative of the enclosed mass with respect to $\rm r$.

By taking the derivative of Eq. \ref{eq:DarkAttr3} with respect to $\rm r$ and equating it to Eq. \ref{eq:Mass_Dens2}, one may directly obtain the halo's density distribution. This results in:
\begin{equation}
\rm \rho_{\rm dark}(r) = \frac{\omega_{0}^{2}}{4\pi G} \cdot ( 3 + 6 \frac{\omega_{\rm bar}(r)}{\omega_{0}} + 2 \frac{\omega_{\rm bar}'(r) \cdot r}{\omega_{0}}) \label{eq:Relation1},
\end{equation}
where $\omega_{\rm bar}'(r)$ is the derivative of $\rm \omega_{\rm bar}(r)$ with respect to $\rm r$.

Eq. \ref{eq:Relation1} may also be formulated in terms of the rotational velocities. This leads to:
\begin{equation}
\rm \rho_{\rm dark}(r) = \frac{\omega_{0}^{2}}{4\pi G}(3 + 4\frac{v_{\rm bar}(r)}{\omega_{0}r} + 2\frac{v'_{\rm bar}(r)}{\omega_{0}}) \label{eq:Relation2},
\end{equation}
where $\rm v_{bar}(r)$ is the baryonic component of a rotation curve and $\rm v_{bar}'(r)$ is the derivative of $\rm v_{bar}(r)$ with respect to $\rm r$.

\section{Rotation Curves behaviour at large radii}\label{AppendixD}

\begin{figure}
\centering
\begin{subfigure}{0.329\textwidth}
    \includegraphics[width=\textwidth]{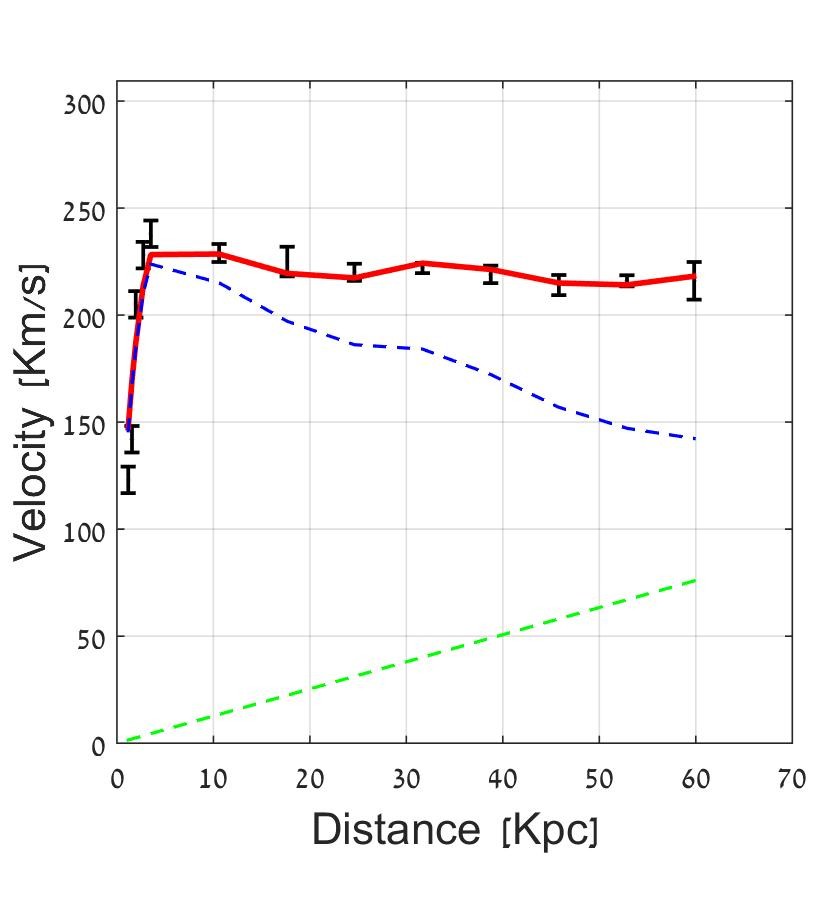}
\end{subfigure}
\begin{subfigure}{0.329\textwidth}
    \includegraphics[width=\textwidth]{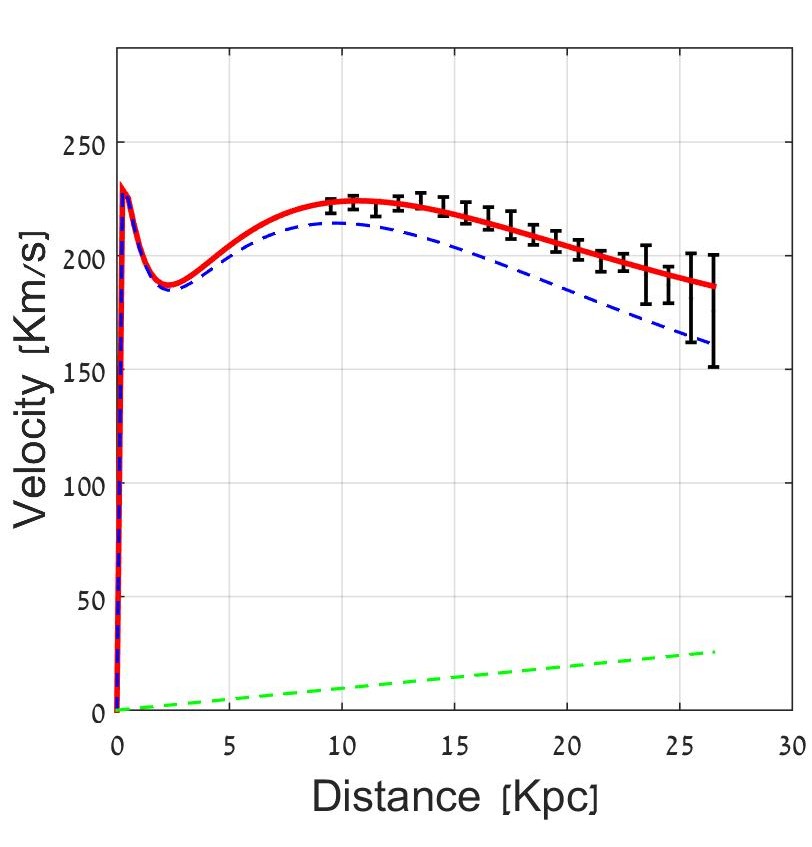}
\end{subfigure}
\begin{subfigure}{0.329\textwidth}
    \includegraphics[width=\textwidth]{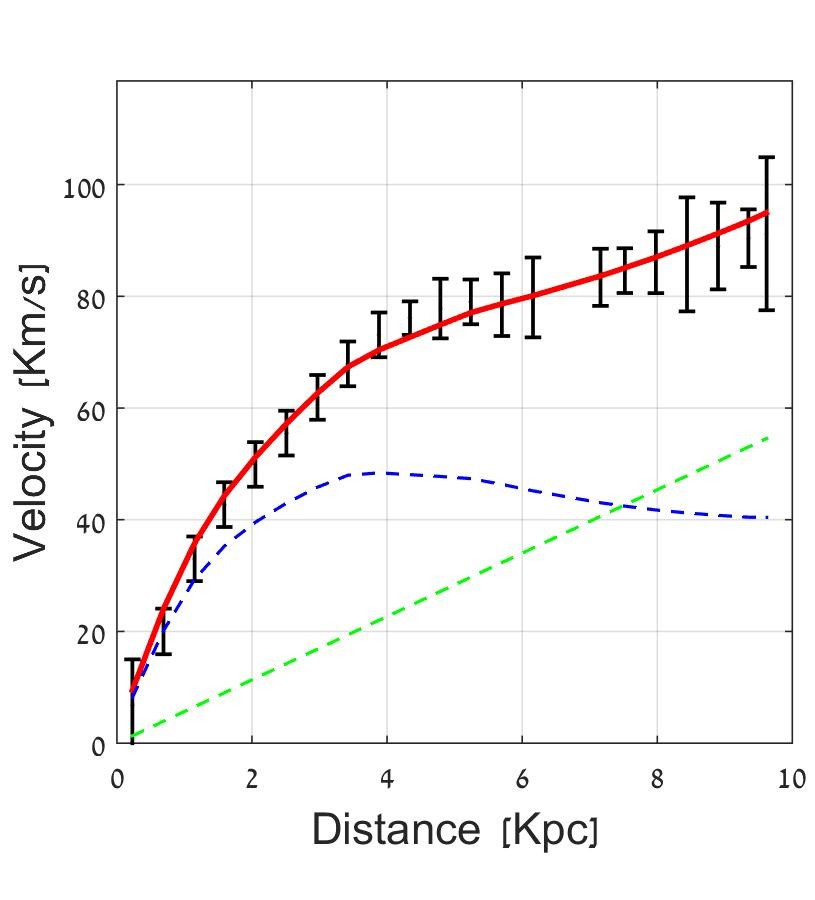}
\end{subfigure}
\caption{Three examples of galactic rotation curves. Top panel: NGC-0801, central panel: Milky-Way, and bottom panel: NGC-0100.
}
    \label{fig:RCs}
\end{figure}

The current study proposes that the difference between an observed AVC and its corresponding baryonic prediction is constant. When adopting this property, a concerning behaviour may arise. In the outer parts of a galaxy, the constant angular velocity term may become dominant, leading to a circular velocity that increases linearly with radius. However, most RCs tend to flatten out at large radii \citep{2001ARAA_39_137S}.

To address this apparent difficulty, it would be beneficial to examine a few examples. Fig.~\ref{fig:RCs} shows three rotation curves, with their data extending to more than four effective radii. The top panel presents a very extended RC (NGC-0801), which demonstrates the model's capability to support long flat behaviour. The best-fit parameters are: $\rm M/L = 0.63 \pm 0.02$ [$\rm M_{\odot} / L_{\odot}$], $\omega_{0} = 0.41 \pm 0.02$ $\rm [10^{-16}\, rad /s]$. The central panel presents the declining RC of the Milky-Way, which demonstrates the model's ability to support galaxies with a dominant baryonic contribution. The data were obtained from a recent study \citep{2023AA_678A_208J_MW_RC}. In this example, an analytical model with a Freeman disk \citep{1970ApJ_Freeman} and a Hernquist bulge \citep{1990ApJ_356_359H_Hernquist_Bulge} was employed. The best-fit parameters are: $\rm M_d = 12.2 \pm 1.2$ $[\rm 10^{10}\, M_{\odot}]$, $\rm R_d = 5.1 \pm 0.3$ $\rm [Kpc]$, $\rm M_b = 1.5 \pm 0.3$ $[\rm 10^{10}\, M_{\odot}]$, $\rm R_b = 0.30 \pm 0.05$ $\rm [Kpc]$ and $\omega_{0} = 0.31 \pm 0.12$ $\rm [10^{-16}\, rad /s]$. Priors for the bulge and disk scale lengths were set according to known values \citep{2023AA_678A_208J_MW_RC}. The bottom panel displays a moderately rising RC (NGC-0100), illustrating the model's ability to support a monotonically increasing behaviour. The best-fit parameters are: $\rm M/L = 0.75 \pm 0.07$ [$\rm M_{\odot} / L_{\odot}$], $\omega_{0} = 1.80 \pm 0.13$ $\rm [10^{-16}\, rad /s]$.

In the first two cases, the rate at which the baryonic term declines is approximately the same as the rate at which the linear term increases over long distances. A visual examination of the entire sample reveals that this kind of behaviour is relatively general. This means that the linear term does not necessarily dictate a linearly rising pattern in the outer parts of an RC. Generally speaking, the model can produce flat, slightly rising, or slightly falling curves in the outer regions of RCs, and the linear term would typically dominate only at very large distances, usually between 2-8 $R_{\rm max}$.

In addition, we remind the data-driven nature of the current research. The model introduced in Eq. \ref{eq:Property} is motivated by its compatibility with rotation curve data. It does not provide a complete theoretical framework that predicts the velocities in any particular situation, e.g. as in \cite{2024ApJ_969L_3M_Flat_Curves}. Future studies may reveal the extent to which this model should be considered valid.

\section{Angular-Velocity Curves}\label{AppendixE}

\begin{figure*}
\centering
\begin{subfigure}{0.33\textwidth}
    \includegraphics[width=\textwidth]{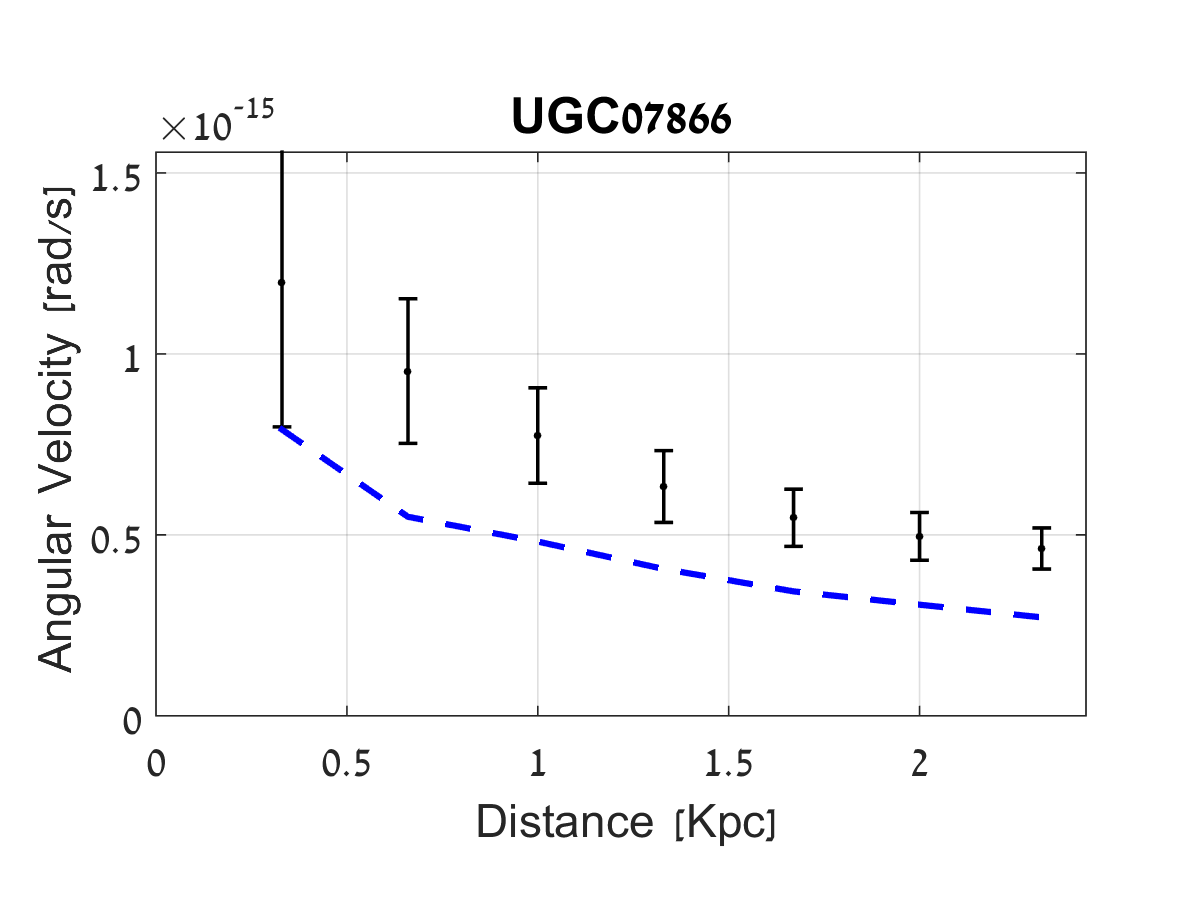}
\end{subfigure}
\begin{subfigure}{0.33\textwidth}
    \includegraphics[width=\textwidth]{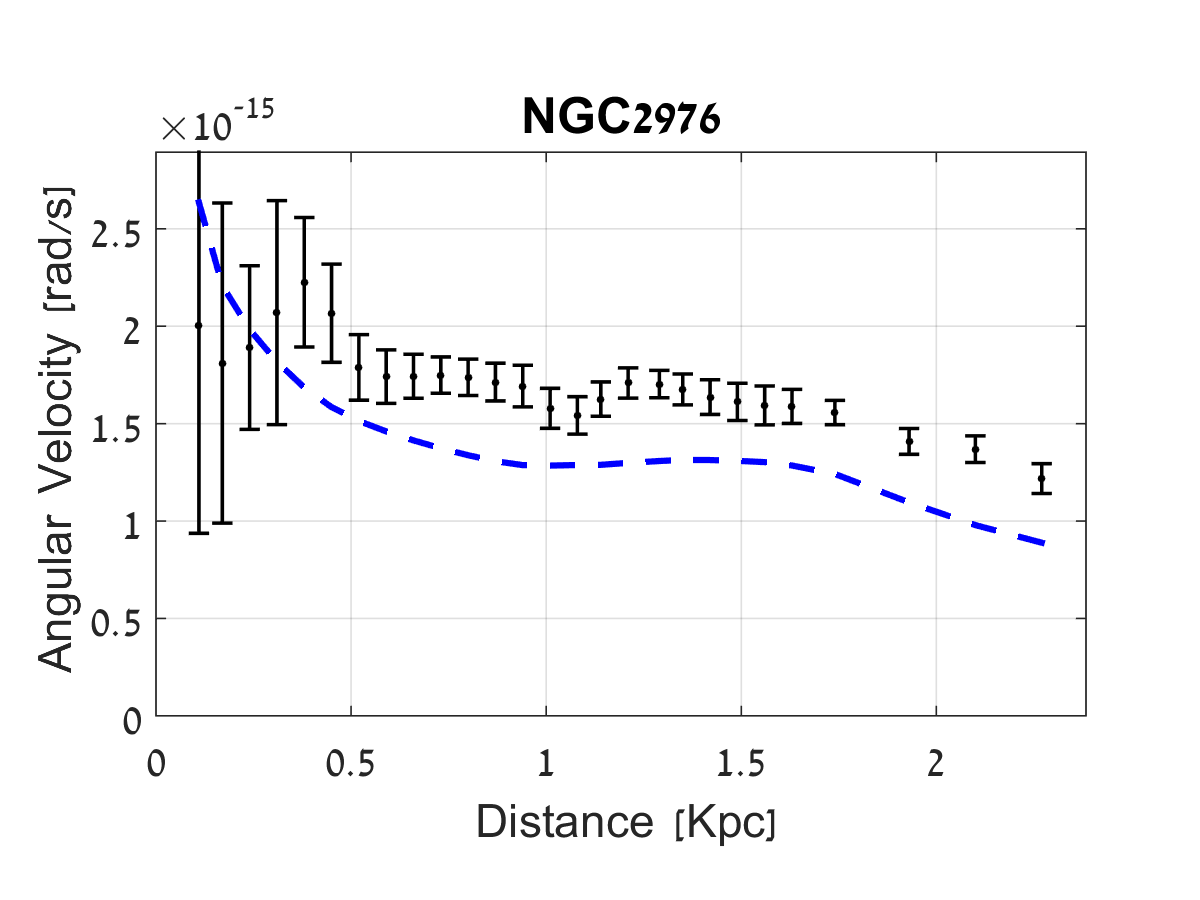}
\end{subfigure}
\begin{subfigure}{0.33\textwidth}
    \includegraphics[width=\textwidth]{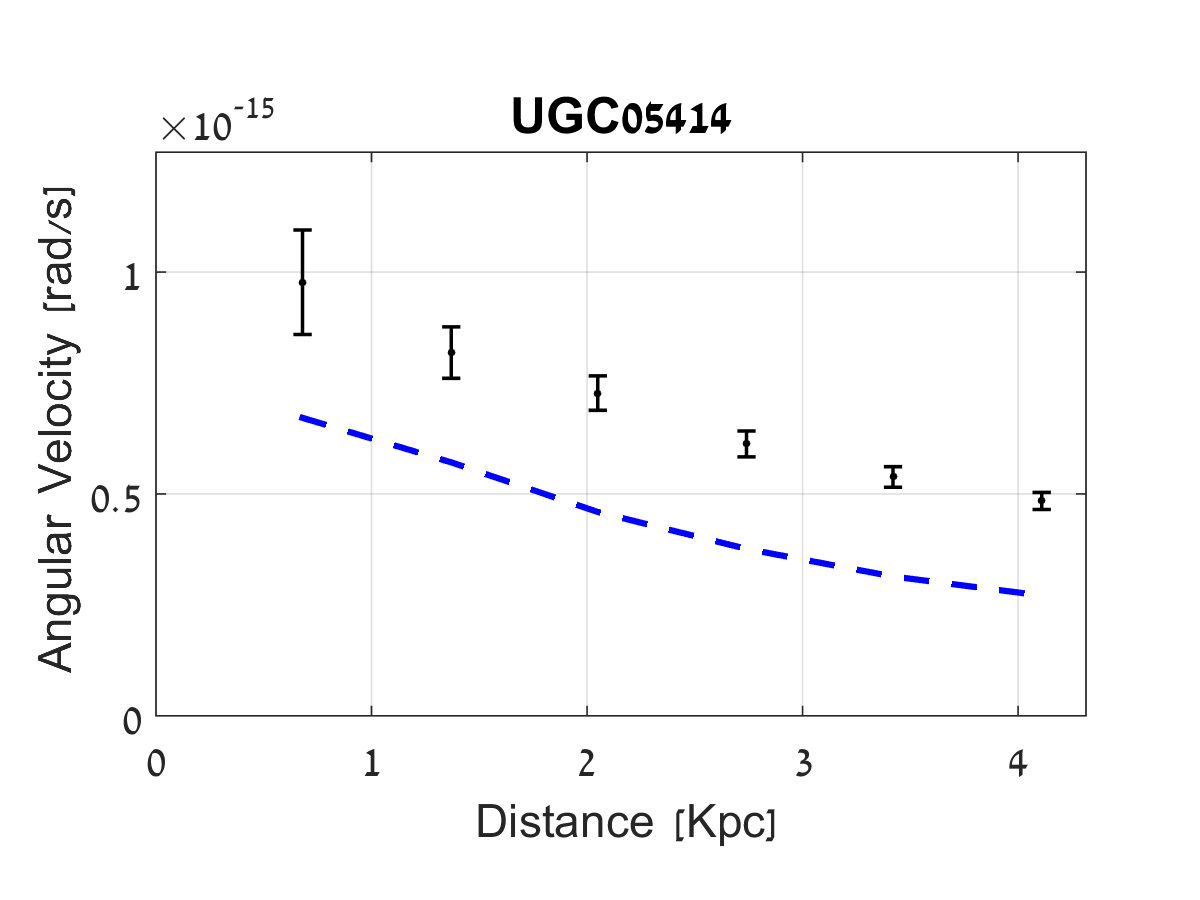}
\end{subfigure}
\begin{subfigure}{0.33\textwidth}
    \includegraphics[width=\textwidth]{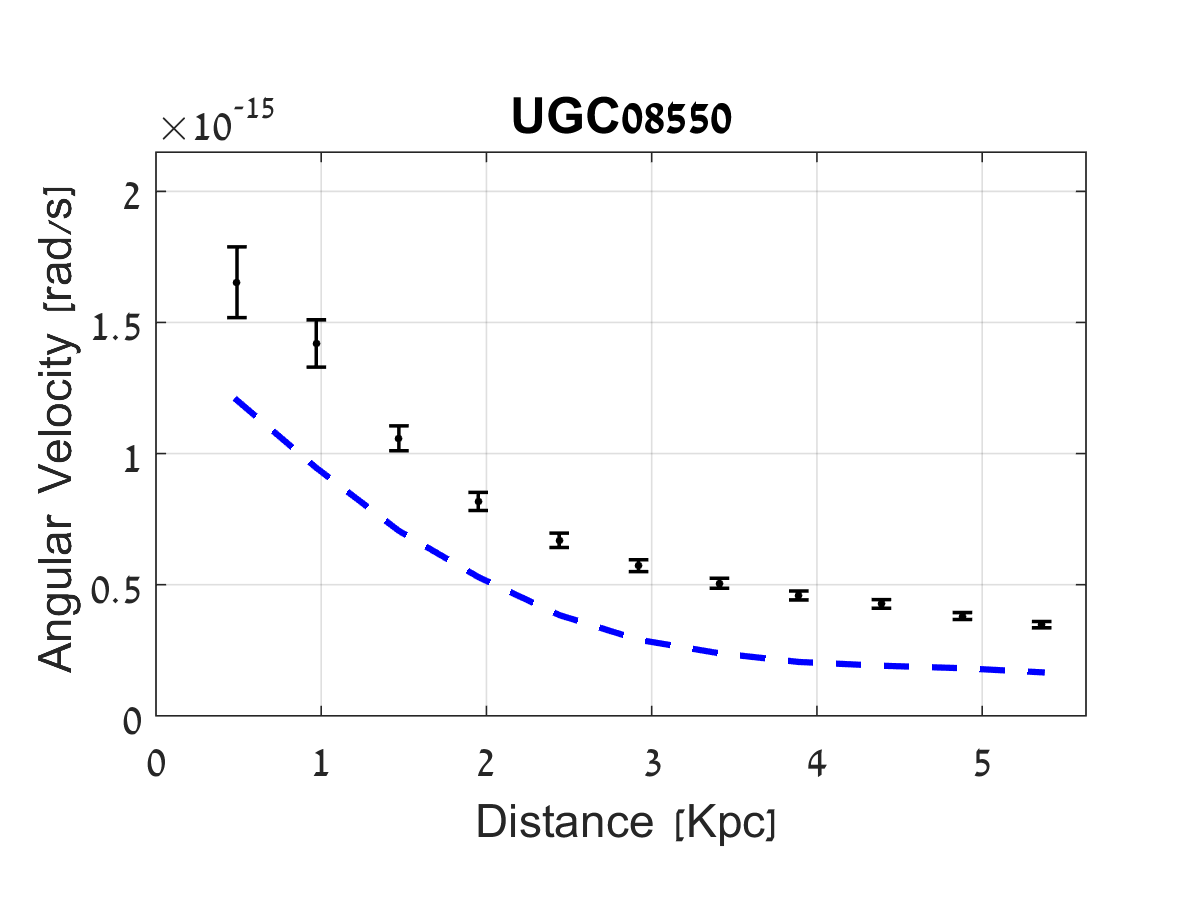}
\end{subfigure}
\begin{subfigure}{0.33\textwidth}
    \includegraphics[width=\textwidth]{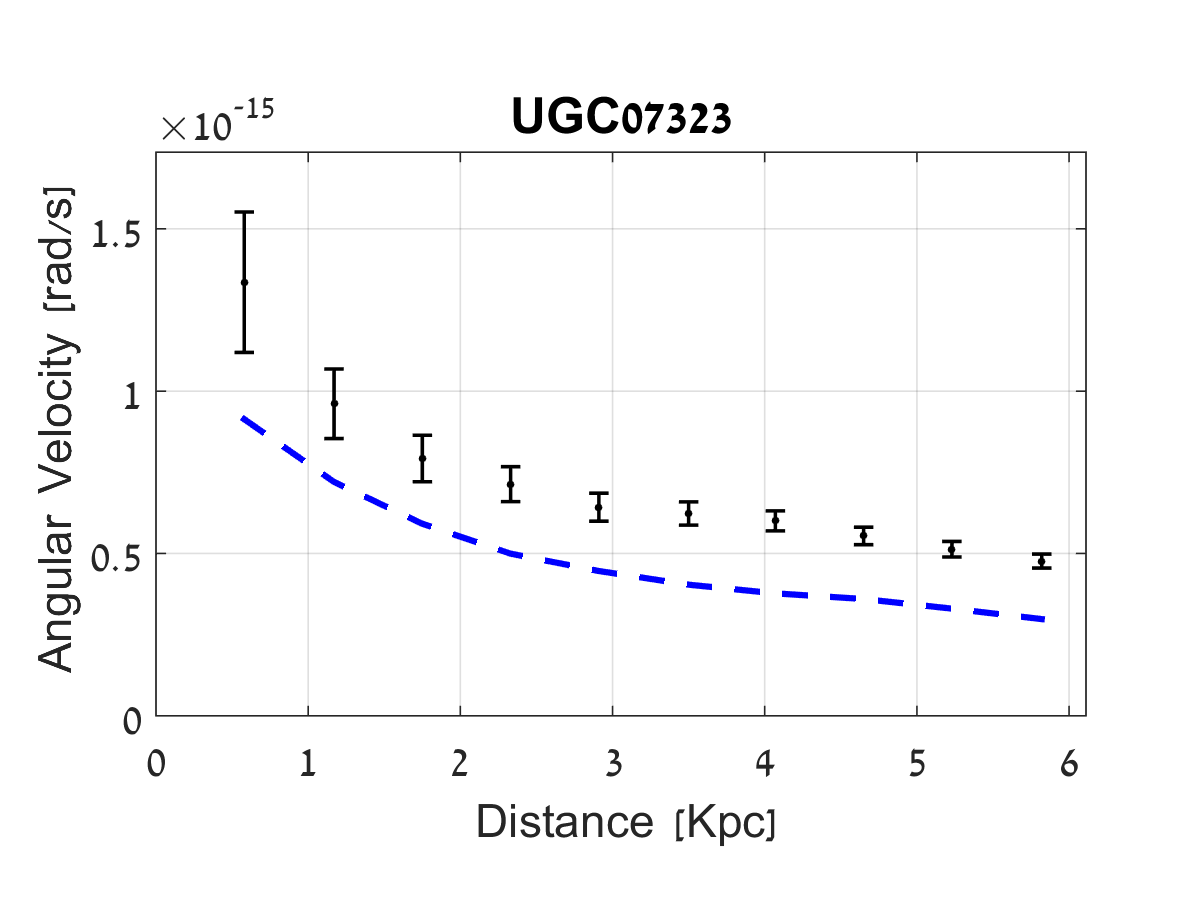}
\end{subfigure}
\begin{subfigure}{0.33\textwidth}
    \includegraphics[width=\textwidth]{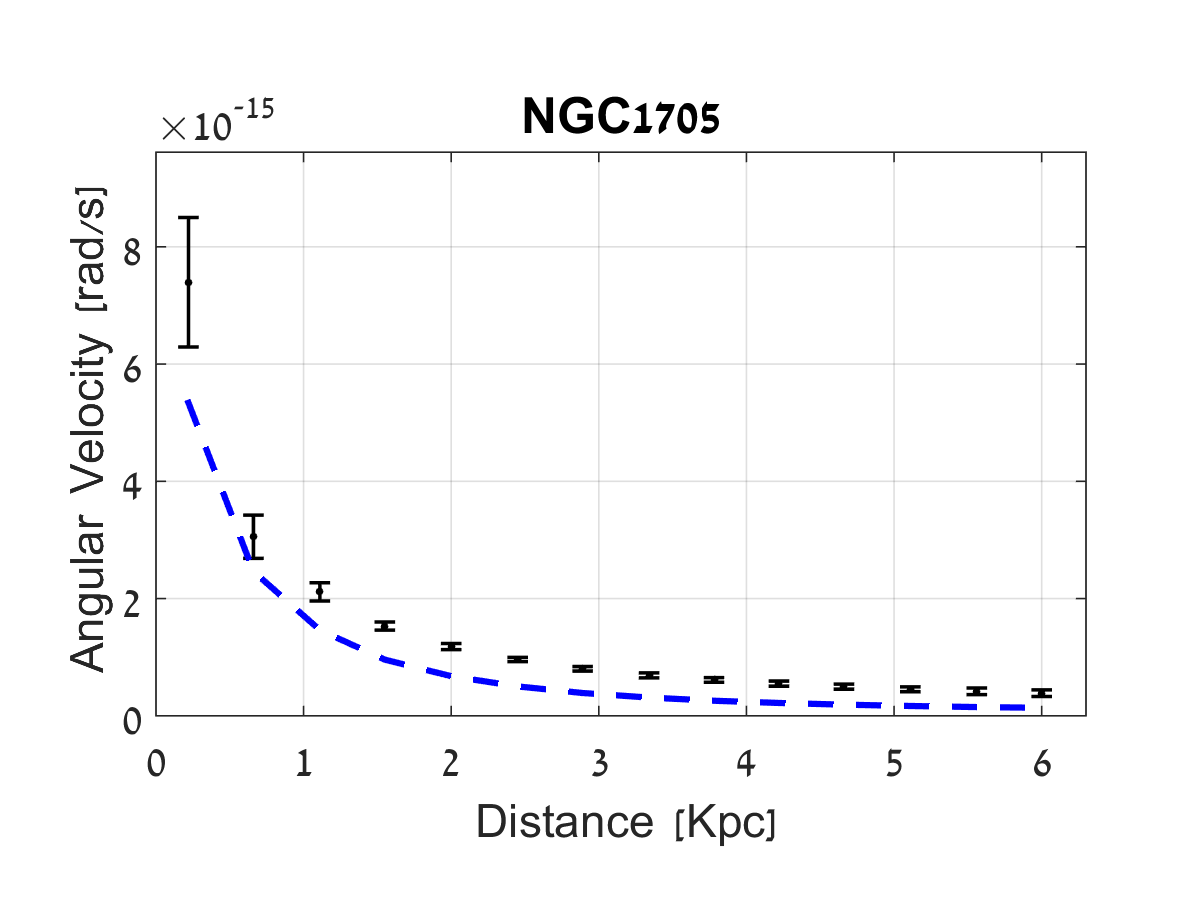}
\end{subfigure}
\begin{subfigure}{0.33\textwidth}
    \includegraphics[width=\textwidth]{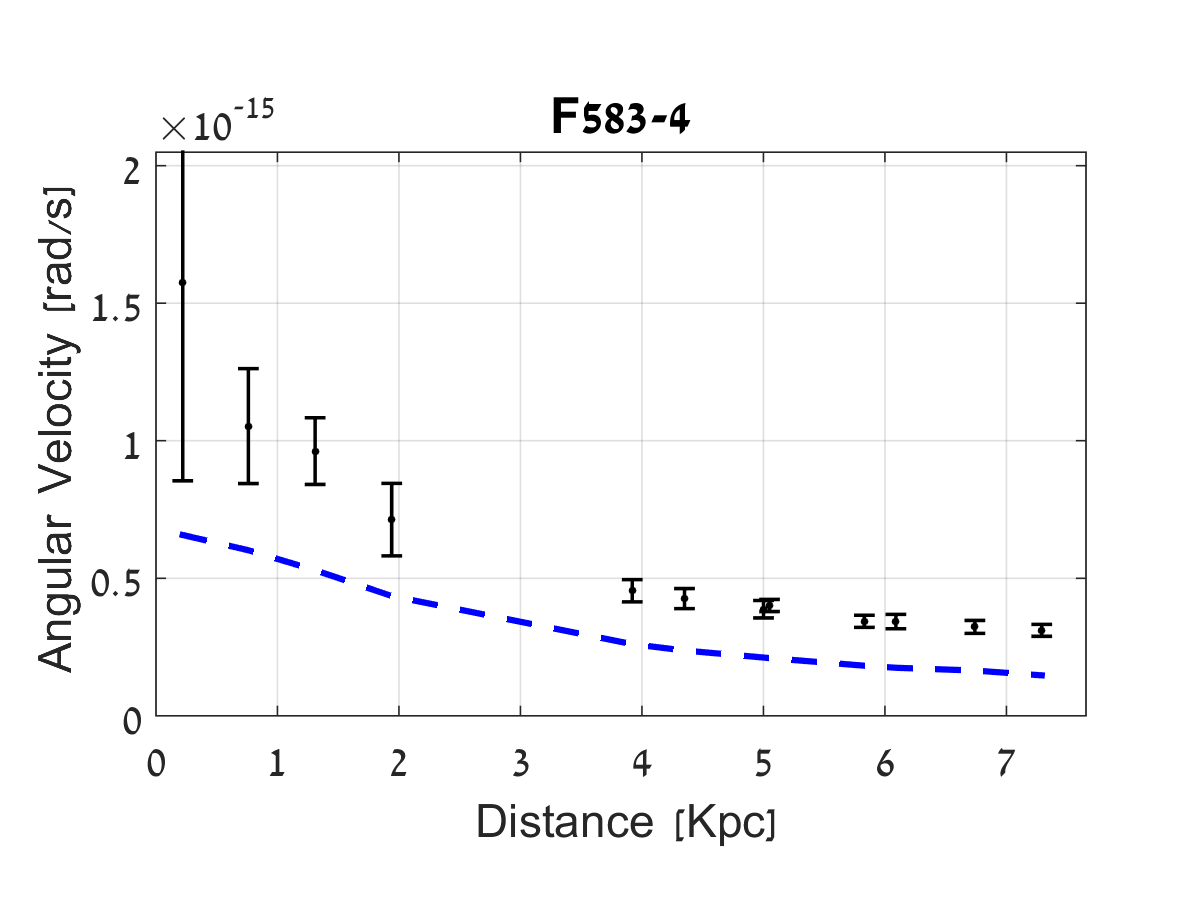}
\end{subfigure}
\begin{subfigure}{0.33\textwidth}
    \includegraphics[width=\textwidth]{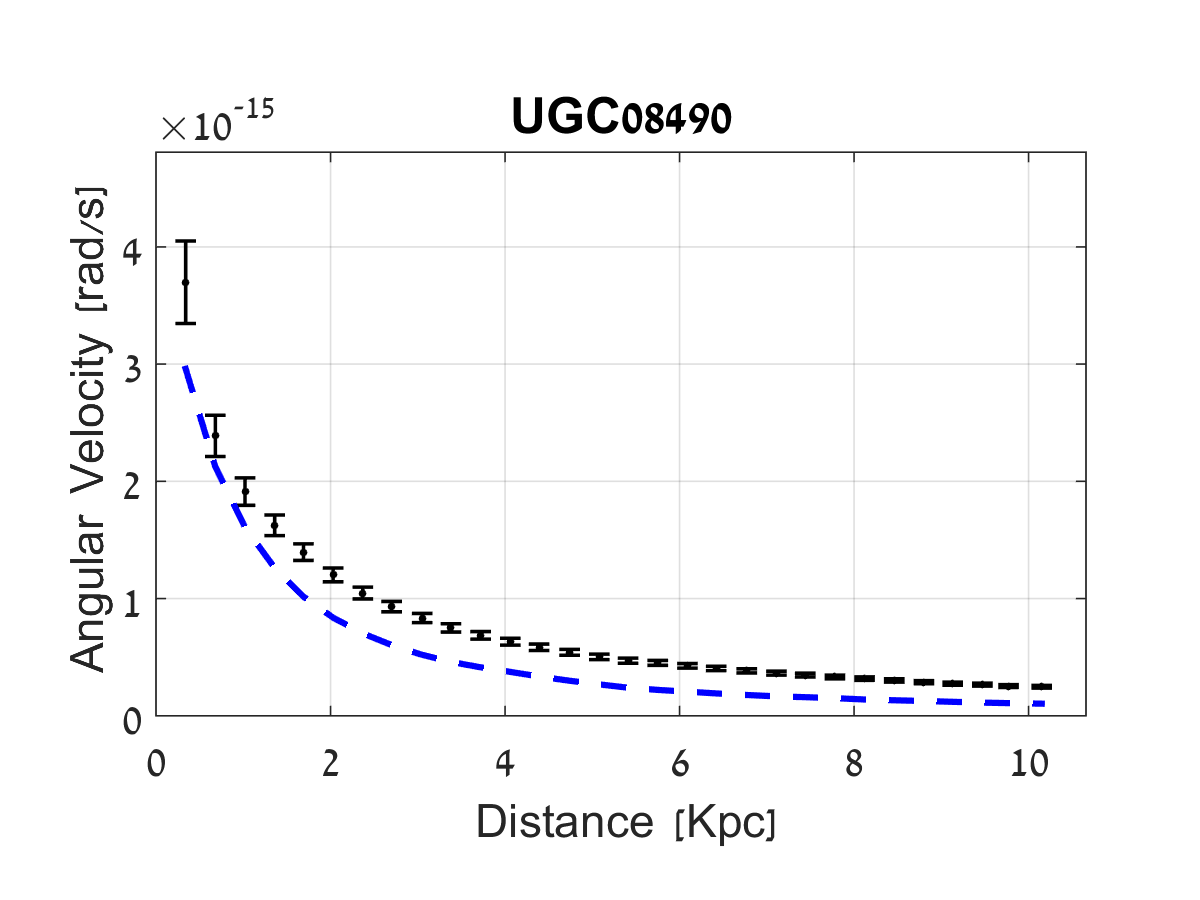}
\end{subfigure}
\begin{subfigure}{0.33\textwidth}
    \includegraphics[width=\textwidth]{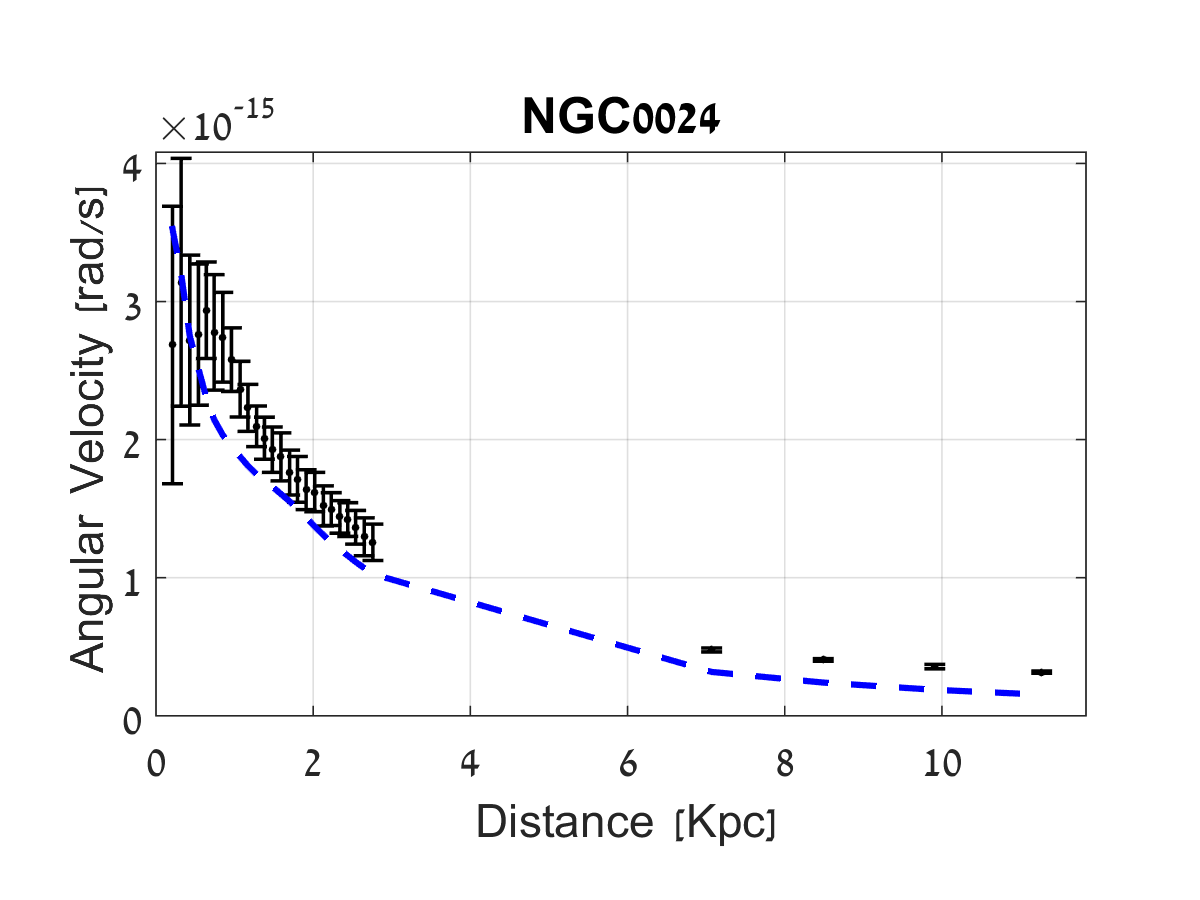}
\end{subfigure}
\begin{subfigure}{0.33\textwidth}
    \includegraphics[width=\textwidth]{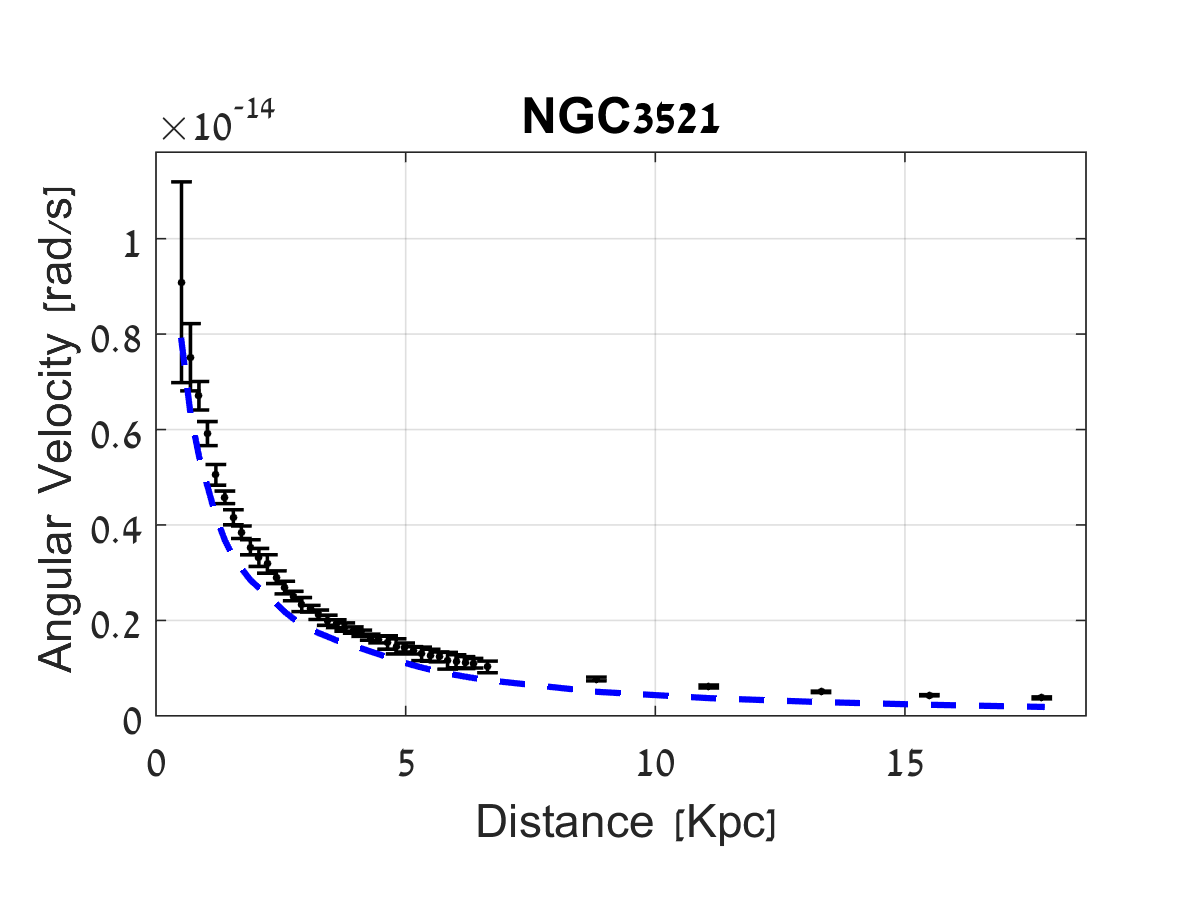}
\end{subfigure}
\begin{subfigure}{0.33\textwidth}
    \includegraphics[width=\textwidth]{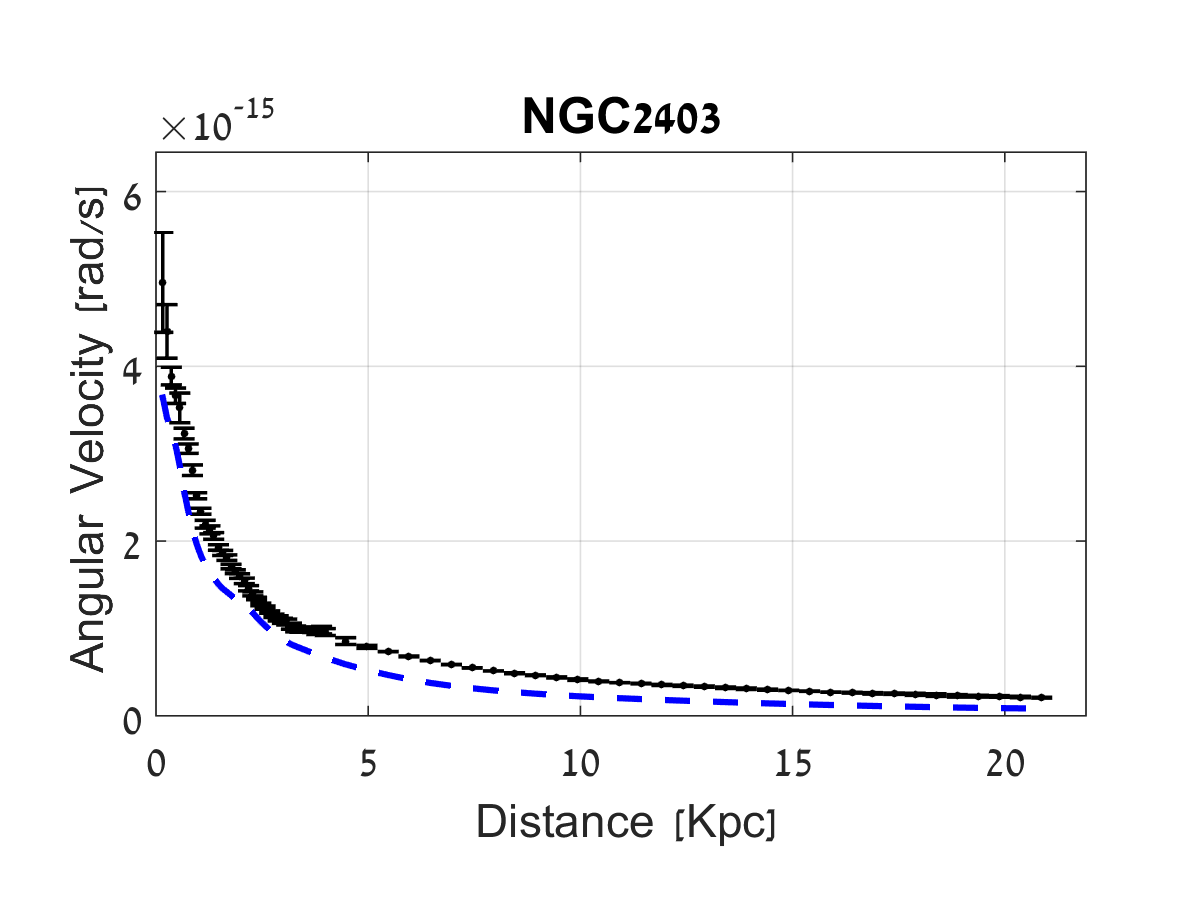}
\end{subfigure}
\begin{subfigure}{0.33\textwidth}
    \includegraphics[width=\textwidth]{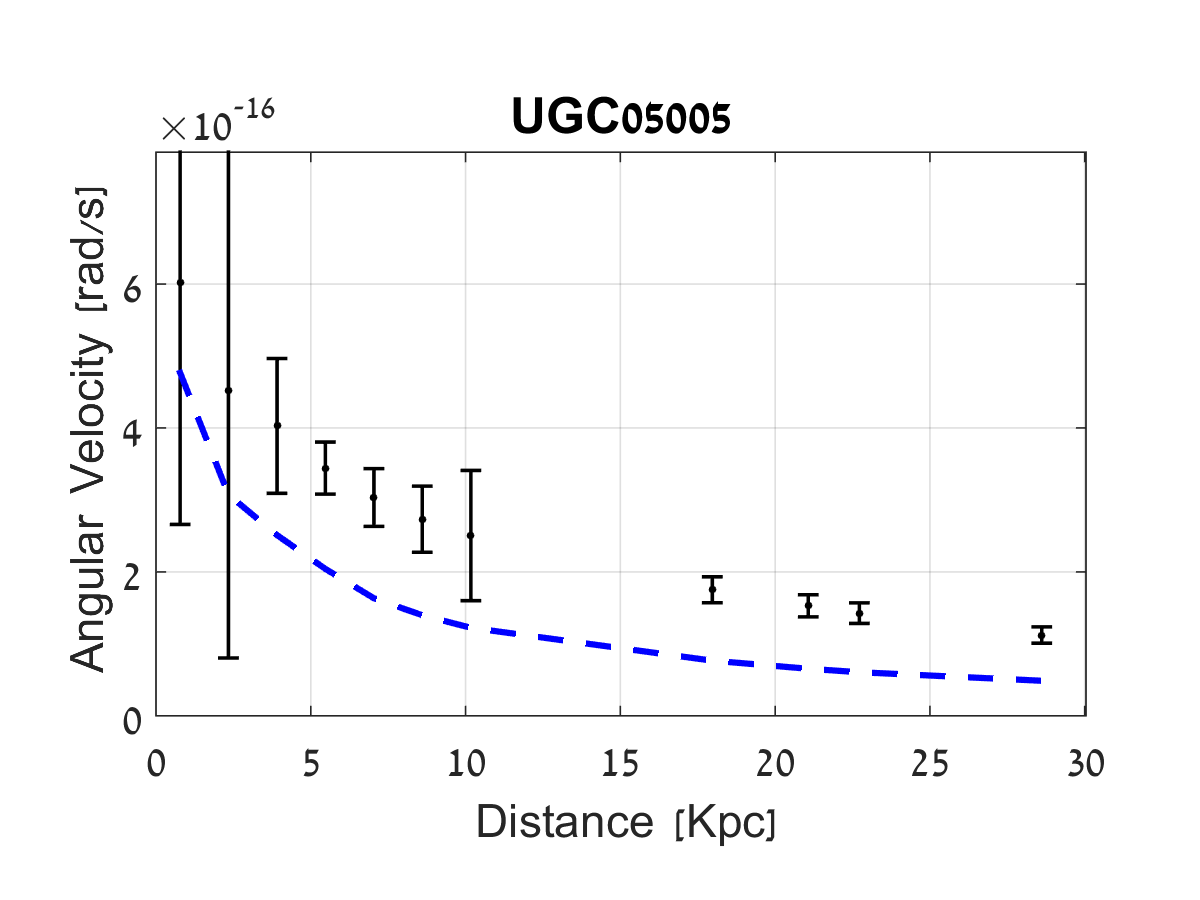}
\end{subfigure}
\caption{
    Measured angular-velocity curves (black points with error-bars) and the baryonic contributions (dashed blue lines). The $M/L$ values: UGC07866: 0.65, NGC2976: 0.55, UGC05414: 0.54, UGC08550: 1.1, UGC07323: 0.55, NGC1705: 1.16, F583-4: 0.63, UGC08490: 1.24, NGC0024: 1.51, NGC3521: 0.4, NGC2403: 0.6, UGC05005: 0.49.
}
    \label{fig:AVC_example2}
\end{figure*}

Fig.~\ref{fig:AVC_example2} presents a selection of angular-velocity curves. These plots are not fits, but rather a visual representations of the data together with the corresponding typical baryonic components. The $\rm M/Ls$ of the baryonic components were taken from \cite{Li_mar_2020}. Plots of this type provided the initial motivation to investigating the universality of constant offsets.

\label{lastpage}
\end{document}